\pdfoutput=1
\documentclass[11pt,twoside,a4paper,pdftex,cmspaper,final,collab]{cms-tdr}

\begin{document}\cmsNoteHeader{CFT-09-023}
%%%%%%%%%%%%%%%%%%%%%%%%%%%%%%%%%%%%%%%%%%%%%%%%%%%%%%%%%%%%%%%%%%%%
%
%  Common definitions
%
%  N.B. use of \providecommand rather than \newcommand means
%       that a definition is ignored if already specified
%
%                                              L. Taylor 18 Feb 2005
%%%%%%%%%%%%%%%%%%%%%%%%%%%%%%%%%%%%%%%%%%%%%%%%%%%%%%%%%%%%%%%%%%%%

%%%%%%%%%%%%%%%%%%%%%%%%%%%%%%%%%%%%%%%%%%%%%%%%%%%%%%%%%%%%%%%%%%%%
%
% Hyphenations (only need to add here if you get a nasty word break)
%
\hyphenation{env-iron-men-tal}%    just an example
\hyphenation{had-ron-i-za-tion}
\hyphenation{cal-or-i-me-ter}
\hyphenation{de-vices}
%
% Hyphenations-end
%
% CVS info. These are modified by cvs at checkout time.
% The last version of these macros found before the maketitle will be the one on the front page,
% so only the main file is tracked.
% Edit by hand with care!
\RCS$Revision: 1.25 $
\RCS$Date: 2010/01/13 15:07:35 $
\RCS$Name:  $
%%%%%%%%%%%%% ptdr definitions %%%%%%%%%%%%%%%%%%%%%
%%%%%%%%%%%%%%%%%%%%%%%%%%%%%%%%%%%%%%%%%%%%%%%%%%%%%%%%%%%%%%%%%%%%
%
%  Common definitions
%
%  N.B. use of \providecommand rather than \newcommand means
%       that a definition is ignored if already specified
%
%                                              L. Taylor 18 Feb 2005
%%%%%%%%%%%%%%%%%%%%%%%%%%%%%%%%%%%%%%%%%%%%%%%%%%%%%%%%%%%%%%%%%%%%

% Some shorthand
% turn off italics
\newcommand {\etal}{\mbox{et al.}\xspace} %et al. - no preceding comma
\newcommand {\ie}{\mbox{i.e.}\xspace}     %i.e.
\newcommand {\eg}{\mbox{e.g.}\xspace}     %e.g.
\newcommand {\etc}{\mbox{etc.}\xspace}     %etc.
\newcommand {\vs}{\mbox{\sl vs.}\xspace}      %vs.
\newcommand {\mdash}{\ensuremath{\mathrm{-}}} % for use within formulas

% some terms whose definition we may change
\newcommand {\Lone}{Level-1\xspace} % Level-1 or L1 ?
\newcommand {\Ltwo}{Level-2\xspace}
\newcommand {\Lthree}{Level-3\xspace}

% Some software programs (alphabetized)
\providecommand{\ACERMC} {\textsc{AcerMC}\xspace}
\providecommand{\ALPGEN} {{\textsc{alpgen}}\xspace}
\providecommand{\CHARYBDIS} {{\textsc{charybdis}}\xspace}
\providecommand{\CMKIN} {\textsc{cmkin}\xspace}
\providecommand{\CMSIM} {{\textsc{cmsim}}\xspace}
\providecommand{\CMSSW} {{\textsc{cmssw}}\xspace}
\providecommand{\COBRA} {{\textsc{cobra}}\xspace}
\providecommand{\COCOA} {{\textsc{cocoa}}\xspace}
\providecommand{\COMPHEP} {\textsc{CompHEP}\xspace}
\providecommand{\EVTGEN} {{\textsc{evtgen}}\xspace}
\providecommand{\FAMOS} {{\textsc{famos}}\xspace}
\providecommand{\GARCON} {\textsc{garcon}\xspace}
\providecommand{\GARFIELD} {{\textsc{garfield}}\xspace}
\providecommand{\GEANE} {{\textsc{geane}}\xspace}
\providecommand{\GEANTfour} {{\textsc{geant4}}\xspace}
\providecommand{\GEANTthree} {{\textsc{geant3}}\xspace}
\providecommand{\GEANT} {{\textsc{geant}}\xspace}
\providecommand{\HDECAY} {\textsc{hdecay}\xspace}
\providecommand{\HERWIG} {{\textsc{herwig}}\xspace}
\providecommand{\HIGLU} {{\textsc{higlu}}\xspace}
\providecommand{\HIJING} {{\textsc{hijing}}\xspace}
\providecommand{\IGUANA} {\textsc{iguana}\xspace}
\providecommand{\ISAJET} {{\textsc{isajet}}\xspace}
\providecommand{\ISAPYTHIA} {{\textsc{isapythia}}\xspace}
\providecommand{\ISASUGRA} {{\textsc{isasugra}}\xspace}
\providecommand{\ISASUSY} {{\textsc{isasusy}}\xspace}
\providecommand{\ISAWIG} {{\textsc{isawig}}\xspace}
\providecommand{\MADGRAPH} {\textsc{MadGraph}\xspace}
\providecommand{\MCATNLO} {\textsc{mc@nlo}\xspace}
\providecommand{\MCFM} {\textsc{mcfm}\xspace}
\providecommand{\MILLEPEDE} {{\textsc{millepede}}\xspace}
\providecommand{\ORCA} {{\textsc{orca}}\xspace}
\providecommand{\OSCAR} {{\textsc{oscar}}\xspace}
\providecommand{\PHOTOS} {\textsc{photos}\xspace}
\providecommand{\PROSPINO} {\textsc{prospino}\xspace}
\providecommand{\PYTHIA} {{\textsc{pythia}}\xspace}
\providecommand{\SHERPA} {{\textsc{sherpa}}\xspace}
\providecommand{\TAUOLA} {\textsc{tauola}\xspace}
\providecommand{\TOPREX} {\textsc{TopReX}\xspace}
\providecommand{\XDAQ} {{\textsc{xdaq}}\xspace}

%  Experiments
\newcommand {\DZERO}{D\O\xspace}     %etc.

% Measurements and units...

\newcommand{\de}{\ensuremath{^\circ}}
\newcommand{\ten}[1]{\ensuremath{\times \text{10}^\text{#1}}}
\newcommand{\unit}[1]{\ensuremath{\text{\,#1}}\xspace}
\newcommand{\mum}{\ensuremath{\,\mu\text{m}}\xspace}
\newcommand{\micron}{\ensuremath{\,\mu\text{m}}\xspace}
\newcommand{\cm}{\ensuremath{\,\text{cm}}\xspace}
\newcommand{\mm}{\ensuremath{\,\text{mm}}\xspace}
\newcommand{\mus}{\ensuremath{\,\mu\text{s}}\xspace}
\newcommand{\keV}{\ensuremath{\,\text{ke\hspace{-.08em}V}}\xspace}
\newcommand{\MeV}{\ensuremath{\,\text{Me\hspace{-.08em}V}}\xspace}
\newcommand{\GeV}{\ensuremath{\,\text{Ge\hspace{-.08em}V}}\xspace}
\newcommand{\TeV}{\ensuremath{\,\text{Te\hspace{-.08em}V}}\xspace}
\newcommand{\PeV}{\ensuremath{\,\text{Pe\hspace{-.08em}V}}\xspace}
\newcommand{\keVc}{\ensuremath{{\,\text{ke\hspace{-.08em}V\hspace{-0.16em}/\hspace{-0.08em}c}}}\xspace}
\newcommand{\MeVc}{\ensuremath{{\,\text{Me\hspace{-.08em}V\hspace{-0.16em}/\hspace{-0.08em}c}}}\xspace}
\newcommand{\GeVc}{\ensuremath{{\,\text{Ge\hspace{-.08em}V\hspace{-0.16em}/\hspace{-0.08em}c}}}\xspace}
\newcommand{\TeVc}{\ensuremath{{\,\text{Te\hspace{-.08em}V\hspace{-0.16em}/\hspace{-0.08em}c}}}\xspace}
\newcommand{\keVcc}{\ensuremath{{\,\text{ke\hspace{-.08em}V\hspace{-0.16em}/\hspace{-0.08em}c}^\text{2}}}\xspace}
\newcommand{\MeVcc}{\ensuremath{{\,\text{Me\hspace{-.08em}V\hspace{-0.16em}/\hspace{-0.08em}c}^\text{2}}}\xspace}
\newcommand{\GeVcc}{\ensuremath{{\,\text{Ge\hspace{-.08em}V\hspace{-0.16em}/\hspace{-0.08em}c}^\text{2}}}\xspace}
\newcommand{\TeVcc}{\ensuremath{{\,\text{Te\hspace{-.08em}V\hspace{-0.16em}/\hspace{-0.08em}c}^\text{2}}}\xspace}

\newcommand{\pbinv} {\mbox{\ensuremath{\,\text{pb}^\text{$-$1}}}\xspace}
\newcommand{\fbinv} {\mbox{\ensuremath{\,\text{fb}^\text{$-$1}}}\xspace}
\newcommand{\nbinv} {\mbox{\ensuremath{\,\text{nb}^\text{$-$1}}}\xspace}
\newcommand{\percms}{\ensuremath{\,\text{cm}^\text{$-$2}\,\text{s}^\text{$-$1}}\xspace}
\newcommand{\lumi}{\ensuremath{\mathcal{L}}\xspace}
\newcommand{\Lumi}{\ensuremath{\mathcal{L}}\xspace}%both upper and lower
%
% Need a convention here:
\newcommand{\LvLow}  {\ensuremath{\mathcal{L}=\text{10}^\text{32}\,\text{cm}^\text{$-$2}\,\text{s}^\text{$-$1}}\xspace}
\newcommand{\LLow}   {\ensuremath{\mathcal{L}=\text{10}^\text{33}\,\text{cm}^\text{$-$2}\,\text{s}^\text{$-$1}}\xspace}
\newcommand{\lowlumi}{\ensuremath{\mathcal{L}=\text{2}\times \text{10}^\text{33}\,\text{cm}^\text{$-$2}\,\text{s}^\text{$-$1}}\xspace}
\newcommand{\LMed}   {\ensuremath{\mathcal{L}=\text{2}\times \text{10}^\text{33}\,\text{cm}^\text{$-$2}\,\text{s}^\text{$-$1}}\xspace}
\newcommand{\LHigh}  {\ensuremath{\mathcal{L}=\text{10}^\text{34}\,\text{cm}^\text{$-$2}\,\text{s}^\text{$-$1}}\xspace}
\newcommand{\hilumi} {\ensuremath{\mathcal{L}=\text{10}^\text{34}\,\text{cm}^\text{$-$2}\,\text{s}^\text{$-$1}}\xspace}

% Some usual physics terms

\newcommand{\zp}{\ensuremath{\mathrm{Z}^\prime}\xspace}

% SM (still to be classified)

\newcommand{\kt}{\ensuremath{k_{\mathrm{T}}}\xspace}
\newcommand{\BC}{\ensuremath{{B_{\mathrm{c}}}}\xspace}
\newcommand{\bbarc}{\ensuremath{{\overline{b}c}}\xspace}
\newcommand{\bbbar}{\ensuremath{{b\overline{b}}}\xspace}
\newcommand{\ccbar}{\ensuremath{{c\overline{c}}}\xspace}
\newcommand{\JPsi}{\ensuremath{{J}/\psi}\xspace}
\newcommand{\bspsiphi}{\ensuremath{B_s \to \JPsi\, \phi}\xspace}
\newcommand{\AFB}{\ensuremath{A_\mathrm{FB}}\xspace}
\newcommand{\EE}{\ensuremath{e^+e^-}\xspace}
\newcommand{\MM}{\ensuremath{\mu^+\mu^-}\xspace}
\newcommand{\TT}{\ensuremath{\tau^+\tau^-}\xspace}
\newcommand{\wangle}{\ensuremath{\sin^{2}\theta_{\mathrm{eff}}^\mathrm{lept}(M^2_\mathrm{Z})}\xspace}
\newcommand{\ttbar}{\ensuremath{{t\overline{t}}}\xspace}
\newcommand{\stat}{\ensuremath{\,\text{(stat.)}}\xspace}
\newcommand{\syst}{\ensuremath{\,\text{(syst.)}}\xspace}
% these moved to similar defs
%\newcommand{\Etmiss}{\ensuremath{E_{\mathrm{T}\!{\rm miss}}}}
%\newcommand{\VEtmiss}{\ensuremath{{\vec E}_{\mathrm{T}\!{\rm miss}}}}

%%%  E-gamma definitions
\newcommand{\HGG}{\ensuremath{\mathrm{H}\to\gamma\gamma}}
\newcommand{\gev}{\GeV}
\newcommand{\GAMJET}{\ensuremath{\gamma + \mathrm{jet}}}
\newcommand{\PPTOJETS}{\ensuremath{\mathrm{pp}\to\mathrm{jets}}}
\newcommand{\PPTOGG}{\ensuremath{\mathrm{pp}\to\gamma\gamma}}
\newcommand{\PPTOGAMJET}{\ensuremath{\mathrm{pp}\to\gamma +
\mathrm{jet}
}}
\newcommand{\MH}{\ensuremath{\mathrm{M_{\mathrm{H}}}}}
\newcommand{\RNINE}{\ensuremath{\mathrm{R}_\mathrm{9}}}
\newcommand{\DR}{\ensuremath{\Delta\mathrm{R}}}

% Physics symbols ...

\newcommand{\PT}{\ensuremath{p_{\mathrm{T}}}\xspace}
\newcommand{\pt}{\ensuremath{p_{\mathrm{T}}}\xspace}
\newcommand{\ET}{\ensuremath{E_{\mathrm{T}}}\xspace}
\newcommand{\HT}{\ensuremath{H_{\mathrm{T}}}\xspace}
\newcommand{\et}{\ensuremath{E_{\mathrm{T}}}\xspace}
\newcommand{\Em}{\ensuremath{E\!\!\!/}\xspace}
\newcommand{\Pm}{\ensuremath{p\!\!\!/}\xspace}
\newcommand{\PTm}{\ensuremath{{p\!\!\!/}_{\mathrm{T}}}\xspace}
\newcommand{\ETm}{\ensuremath{E_{\mathrm{T}}^{\mathrm{miss}}}\xspace}
\newcommand{\MET}{\ensuremath{E_{\mathrm{T}}^{\mathrm{miss}}}\xspace}
\newcommand{\ETmiss}{\ensuremath{E_{\mathrm{T}}^{\mathrm{miss}}}\xspace}
\newcommand{\VEtmiss}{\ensuremath{{\vec E}_{\mathrm{T}}^{\mathrm{miss}}}\xspace}

%%%%%%
% From Albert
%

\newcommand{\ga}{\ensuremath{\gtrsim}}
\newcommand{\la}{\ensuremath{\lesssim}}
\newcommand{\swsq}{\ensuremath{\sin^2\theta_W}\xspace}
\newcommand{\cwsq}{\ensuremath{\cos^2\theta_W}\xspace}
\newcommand{\tanb}{\ensuremath{\tan\beta}\xspace}
\newcommand{\tanbsq}{\ensuremath{\tan^{2}\beta}\xspace}
\newcommand{\sidb}{\ensuremath{\sin 2\beta}\xspace}
\newcommand{\alpS}{\ensuremath{\alpha_S}\xspace}
\newcommand{\alpt}{\ensuremath{\tilde{\alpha}}\xspace}

\newcommand{\QL}{\ensuremath{Q_L}\xspace}
\newcommand{\sQ}{\ensuremath{\tilde{Q}}\xspace}
\newcommand{\sQL}{\ensuremath{\tilde{Q}_L}\xspace}
\newcommand{\ULC}{\ensuremath{U_L^C}\xspace}
\newcommand{\sUC}{\ensuremath{\tilde{U}^C}\xspace}
\newcommand{\sULC}{\ensuremath{\tilde{U}_L^C}\xspace}
\newcommand{\DLC}{\ensuremath{D_L^C}\xspace}
\newcommand{\sDC}{\ensuremath{\tilde{D}^C}\xspace}
\newcommand{\sDLC}{\ensuremath{\tilde{D}_L^C}\xspace}
\newcommand{\LL}{\ensuremath{L_L}\xspace}
\newcommand{\sL}{\ensuremath{\tilde{L}}\xspace}
\newcommand{\sLL}{\ensuremath{\tilde{L}_L}\xspace}
\newcommand{\ELC}{\ensuremath{E_L^C}\xspace}
\newcommand{\sEC}{\ensuremath{\tilde{E}^C}\xspace}
\newcommand{\sELC}{\ensuremath{\tilde{E}_L^C}\xspace}
\newcommand{\sEL}{\ensuremath{\tilde{E}_L}\xspace}
\newcommand{\sER}{\ensuremath{\tilde{E}_R}\xspace}
\newcommand{\sFer}{\ensuremath{\tilde{f}}\xspace}
\newcommand{\sQua}{\ensuremath{\tilde{q}}\xspace}
\newcommand{\sUp}{\ensuremath{\tilde{u}}\xspace}
\newcommand{\suL}{\ensuremath{\tilde{u}_L}\xspace}
\newcommand{\suR}{\ensuremath{\tilde{u}_R}\xspace}
\newcommand{\sDw}{\ensuremath{\tilde{d}}\xspace}
\newcommand{\sdL}{\ensuremath{\tilde{d}_L}\xspace}
\newcommand{\sdR}{\ensuremath{\tilde{d}_R}\xspace}
\newcommand{\sTop}{\ensuremath{\tilde{t}}\xspace}
\newcommand{\stL}{\ensuremath{\tilde{t}_L}\xspace}
\newcommand{\stR}{\ensuremath{\tilde{t}_R}\xspace}
\newcommand{\stone}{\ensuremath{\tilde{t}_1}\xspace}
\newcommand{\sttwo}{\ensuremath{\tilde{t}_2}\xspace}
\newcommand{\sBot}{\ensuremath{\tilde{b}}\xspace}
\newcommand{\sbL}{\ensuremath{\tilde{b}_L}\xspace}
\newcommand{\sbR}{\ensuremath{\tilde{b}_R}\xspace}
\newcommand{\sbone}{\ensuremath{\tilde{b}_1}\xspace}
\newcommand{\sbtwo}{\ensuremath{\tilde{b}_2}\xspace}
\newcommand{\sLep}{\ensuremath{\tilde{l}}\xspace}
\newcommand{\sLepC}{\ensuremath{\tilde{l}^C}\xspace}
\newcommand{\sEl}{\ensuremath{\tilde{e}}\xspace}
\newcommand{\sElC}{\ensuremath{\tilde{e}^C}\xspace}
\newcommand{\seL}{\ensuremath{\tilde{e}_L}\xspace}
\newcommand{\seR}{\ensuremath{\tilde{e}_R}\xspace}
\newcommand{\snL}{\ensuremath{\tilde{\nu}_L}\xspace}
\newcommand{\sMu}{\ensuremath{\tilde{\mu}}\xspace}
\newcommand{\sNu}{\ensuremath{\tilde{\nu}}\xspace}
\newcommand{\sTau}{\ensuremath{\tilde{\tau}}\xspace}
\newcommand{\Glu}{\ensuremath{g}\xspace}
\newcommand{\sGlu}{\ensuremath{\tilde{g}}\xspace}
\newcommand{\Wpm}{\ensuremath{W^{\pm}}\xspace}
\newcommand{\sWpm}{\ensuremath{\tilde{W}^{\pm}}\xspace}
\newcommand{\Wz}{\ensuremath{W^{0}}\xspace}
\newcommand{\sWz}{\ensuremath{\tilde{W}^{0}}\xspace}
\newcommand{\sWino}{\ensuremath{\tilde{W}}\xspace}
\newcommand{\Bz}{\ensuremath{B^{0}}\xspace}
\newcommand{\sBz}{\ensuremath{\tilde{B}^{0}}\xspace}
\newcommand{\sBino}{\ensuremath{\tilde{B}}\xspace}
\newcommand{\Zz}{\ensuremath{Z^{0}}\xspace}
\newcommand{\sZino}{\ensuremath{\tilde{Z}^{0}}\xspace}
\newcommand{\sGam}{\ensuremath{\tilde{\gamma}}\xspace}
\newcommand{\chiz}{\ensuremath{\tilde{\chi}^{0}}\xspace}
\newcommand{\chip}{\ensuremath{\tilde{\chi}^{+}}\xspace}
\newcommand{\chim}{\ensuremath{\tilde{\chi}^{-}}\xspace}
\newcommand{\chipm}{\ensuremath{\tilde{\chi}^{\pm}}\xspace}
\newcommand{\Hone}{\ensuremath{H_{d}}\xspace}
\newcommand{\sHone}{\ensuremath{\tilde{H}_{d}}\xspace}
\newcommand{\Htwo}{\ensuremath{H_{u}}\xspace}
\newcommand{\sHtwo}{\ensuremath{\tilde{H}_{u}}\xspace}
\newcommand{\sHig}{\ensuremath{\tilde{H}}\xspace}
\newcommand{\sHa}{\ensuremath{\tilde{H}_{a}}\xspace}
\newcommand{\sHb}{\ensuremath{\tilde{H}_{b}}\xspace}
\newcommand{\sHpm}{\ensuremath{\tilde{H}^{\pm}}\xspace}
\newcommand{\hz}{\ensuremath{h^{0}}\xspace}
\newcommand{\Hz}{\ensuremath{H^{0}}\xspace}
\newcommand{\Az}{\ensuremath{A^{0}}\xspace}
\newcommand{\Hpm}{\ensuremath{H^{\pm}}\xspace}
\newcommand{\sGra}{\ensuremath{\tilde{G}}\xspace}
\newcommand{\mtil}{\ensuremath{\tilde{m}}\xspace}
\newcommand{\rpv}{\ensuremath{\rlap{\kern.2em/}R}\xspace}
\newcommand{\LLE}{\ensuremath{LL\bar{E}}\xspace}
\newcommand{\LQD}{\ensuremath{LQ\bar{D}}\xspace}
\newcommand{\UDD}{\ensuremath{\overline{UDD}}\xspace}
\newcommand{\Lam}{\ensuremath{\lambda}\xspace}
\newcommand{\Lamp}{\ensuremath{\lambda'}\xspace}
\newcommand{\Lampp}{\ensuremath{\lambda''}\xspace}
\newcommand{\spinbd}[2]{\ensuremath{\bar{#1}_{\dot{#2}}}\xspace}

\newcommand{\MD}{\ensuremath{{M_\mathrm{D}}}\xspace}% ED mass
\newcommand{\Mpl}{\ensuremath{{M_\mathrm{Pl}}}\xspace}% Planck mass
\newcommand{\Rinv} {\ensuremath{{R}^{-1}}\xspace}

%%%%%%%%%%%%%%%%%%%%%%%%%%%%%%%%%%%%%%%%%%%%%%%%%%%%%%%%%%%%%%%%%%%%
%
% Hyphenations (only need to add here if you get a nasty word break)
%
\hyphenation{en-viron-men-tal}%    just an example

%%%%%%%%%%%%%%%  Title page %%%%%%%%%%%%%%%%%%%%%%%%
%\title{Calibration of the CMS Drift-Tube System and Drift Velocity Measurements with Cosmic Muons }

\cmsNoteHeader{09-023}
\title{Calibration of the CMS Drift Tube Chambers and Measurement of the Drift Velocity with Cosmic Rays}

%Author is always "The CMS Collaboration" for PAS, so author, etc will be ignored
%\address[cern]{CERN}
%\address[neu]{Northeastern University}
%\author[neu]{George Alverson}\author[neu]{Lucas Taylor}\author[cern]{A. Cern Person}

% please supply the date in yyyy/mm/dd format. Today has been
% redefined to do so, but it should be fixed as of the final release date.
\date{\today}

% note that you cannot use \verb in the abstract text
\abstract{ This paper describes the calibration procedure for the
  drift tubes of the CMS barrel muon system and reports the main
  results obtained with data collected during a high statistics
  cosmic ray data-taking period.
  The main goal of the calibration is to determine, for each drift
  cell, the minimum time delay for signals relative to the trigger,
  accounting for
  the drift velocity within the cell. The accuracy of the calibration
  procedure is influenced by the random arrival time of the cosmic
  muons relative to the LHC clock cycle.
%  The accuracy of the calibration
%  procedure is influenced by the random arrival time of cosmic muons.
  A more refined analysis of the drift velocity was performed during
  the offline reconstruction phase, which takes into account this
  feature of cosmic ray events.  }

% these need to be filled in by hand and should (MUST) match the info
% in the TeX equivalents less the TeX markup
\hypersetup{%
pdfauthor={Silvia Maselli},%
pdftitle={Calibration of the CMS Drift Tube Chambers and Measurement of the Drift Velocity with Cosmic Rays},%
pdfsubject={CMS DT Calibration},%
pdfkeywords={CMS, DT, software, calibration, CRAFT}}

\maketitle %maketitle comes after all the front information has been supplied

\section{Introduction}

The Compact Muon Solenoid (CMS) \cite{CMS} is a general-purpose
detector whose main goal is to explore physics at the TeV scale, by
exploiting the proton-proton collisions provided by the Large Hadron
Collider (LHC) \cite{LHC} at CERN.

CMS uses a
right-handed coordinate system, with the origin at the nominal
collision point, the $x$-axis pointing to the center of the LHC, the
$y$-axis pointing up (perpendicular to the LHC plane), and the
$z$-axis along the anticlockwise-beam direction. The polar angle,
$\theta$, is measured from the positive $z$-axis and the azimuthal
angle, $\phi$, is measured in the $x$-$y$ plane.

The central feature of the Compact Muon Solenoid apparatus is a
superconducting solenoid, of $6~\rm m$ internal diameter, providing a field of
$3.8~\rm T$. Within the field volume are the silicon pixel and strip tracker, the
crystal electromagnetic calorimeter (ECAL) and the brass/scintillator hadron
calorimeter (HCAL).  Muons are measured in gas-ionization detectors embedded
in the steel return yoke.  In addition to the barrel and endcap detectors,
CMS has extensive forward calorimetry.

\begin{figure}[htbp]
 \begin{center}
   \resizebox{10cm}{!}{\includegraphics{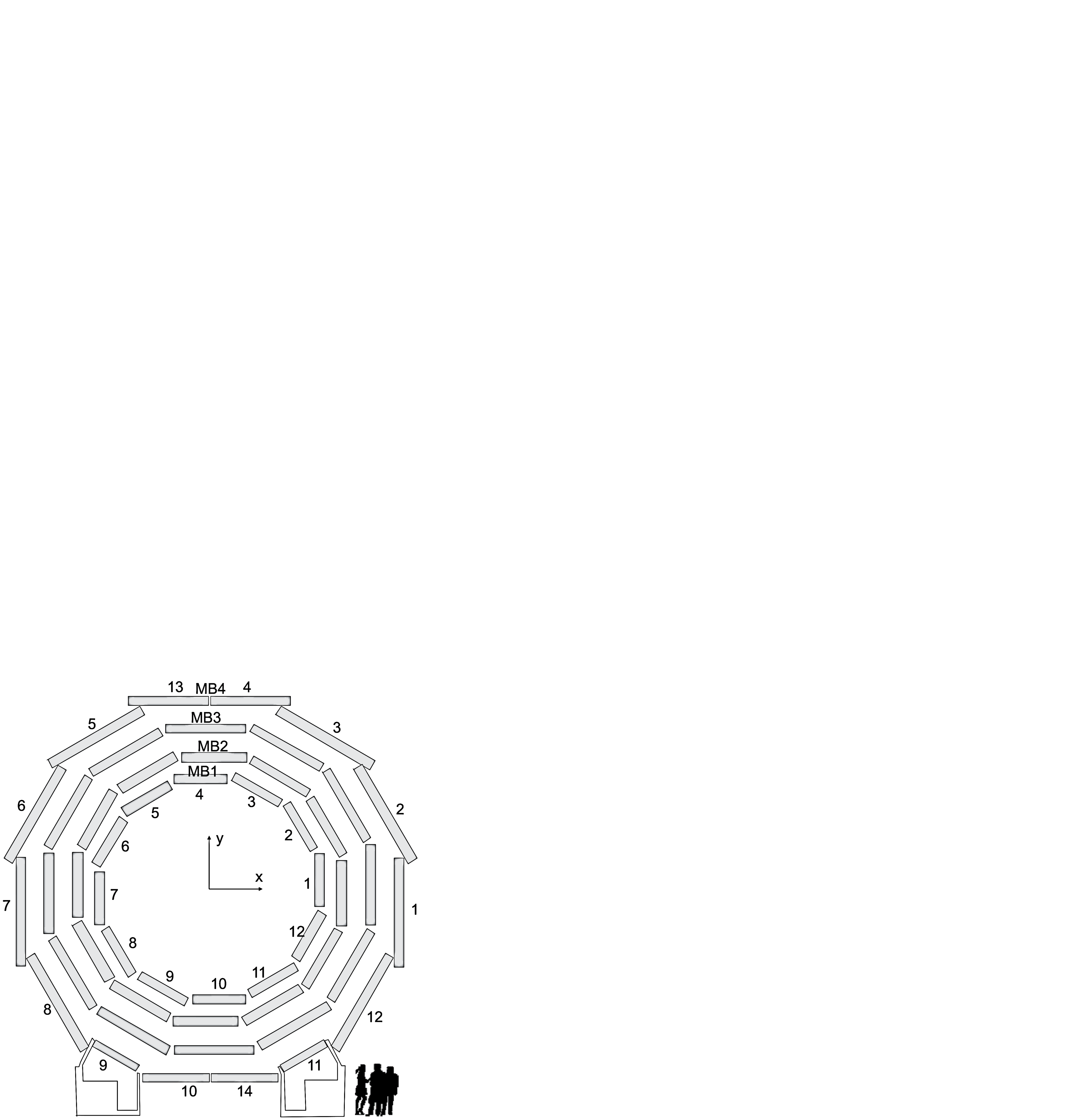}}
  \caption{Schematic representation, in the $x-y$
   plane, of the chamber positions within a
   wheel of the muon barrel system of the CMS experiment.  
   The labels and the numbers of the muon stations are shown. Because of
   mechanical requirements, the top and bottom MB4 sectors are 
 split in two distinct chambers.}
  \label{fig:CMS}
 \end{center}
\end{figure}

The barrel muon system \cite{MUON} is divided in five wheels. Every
wheel is composed of 12 sectors, each covering 
30$^\circ$ in azimuth, as shown in Fig.~\ref{fig:CMS}.  

\begin{figure}[htbp]
 \begin{center}
  \begin{tabular}{c}

   \resizebox{8cm}{!}{\includegraphics{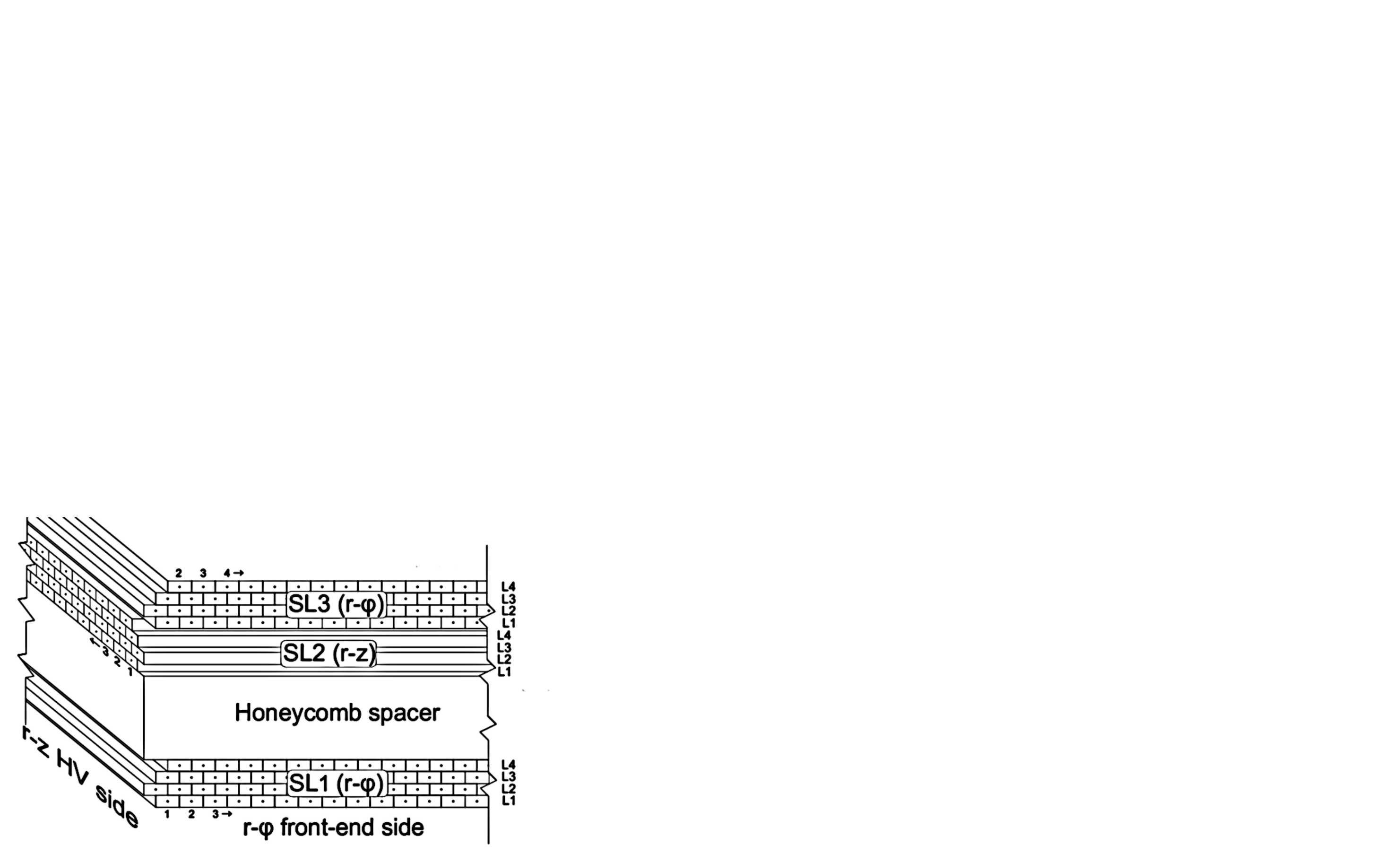}}
   \resizebox{8cm}{!}{\includegraphics{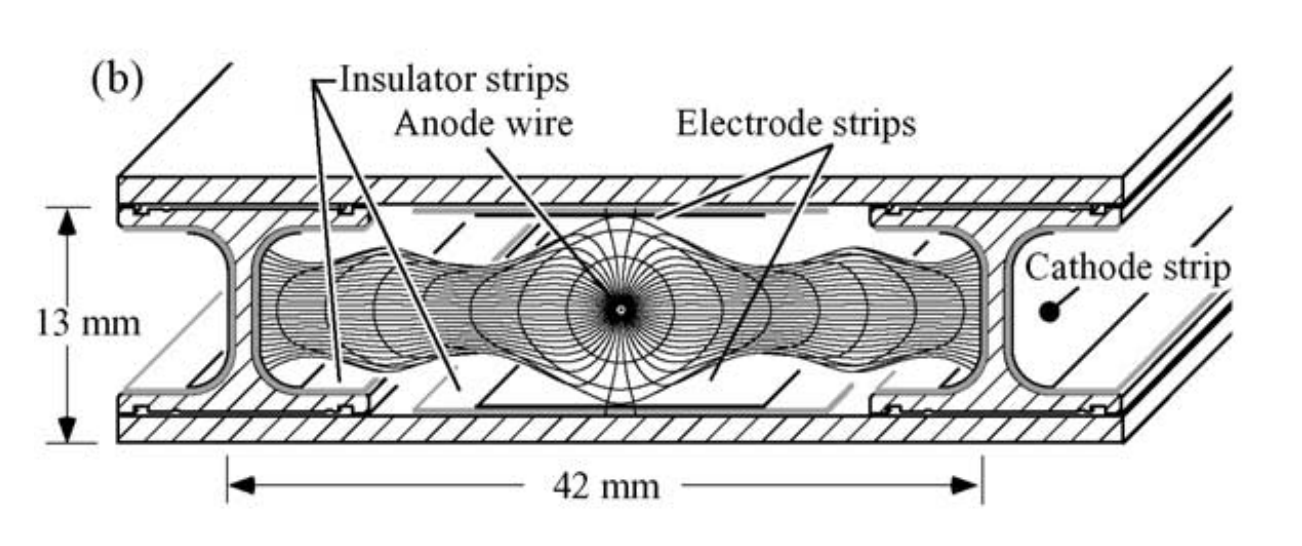}}
 \end{tabular}

 \caption{Left: Schematic view of a DT chamber. Right: Section of a
   drift tube cell showing drift lines and isochrones. The voltages
   applied are $+3600~\rm V$ for wires, $+1800~\rm V$ for electrode
   strips, and $-1200~\rm V$ for cathode strips. }
  \label{fig:DTChamCell}
 \end{center}
\end{figure}

Each sector contains four stations equipped with Resistive Plate
Chambers (RPC) and Drift-Tubes (DT) chambers.  The four DT chambers
are labeled MB1, MB2, MB3, and MB4 going inside-out.
In total there are $5~\rm wheels * ( 3~\rm stations * 12~\rm sectors + 1~\rm station * 14~\rm sectors) = 250 $ DT chambers. 
The chambers are interleaved with the steel return yoke
of the magnet and are composed of three groups, called
``super-layers'' (SL), of four staggered layers
of independent drift cells, for a total of
about 172\,000 channels.  A schematic representation of a chamber is
shown in Fig.~\ref{fig:DTChamCell} (left).

The chamber volume is filled with a Ar(85\,\%)/CO$_2$(15\,\%) gas
mixture, kept at atmospheric pressure.  Two of the super-layers have
the wires parallel to the beam direction and measure the $r\phi$
coordinate, the other super-layer has wires perpendicular to the beam
direction and measures the $z$ coordinate. 
The chamber provides a measurement of a track segment in space. 
The outermost station is
equipped with chambers containing only the two $r\phi$
super-layers. The basic element of the DT detector is the drift cell,
 illustrated in Fig.~\ref{fig:DTChamCell} (right), where
the drift lines and isochrones are represented.
All chambers were operational, fully commissioned, 
and the number of problematic channels were less than $1~\%$.
% in several cosmic ray data-taking periods.

The DT system is designed to provide muon track reconstruction, with
the correct charge assignment up to TeV energies, and
first-level trigger selection. It also yields a
fast muon identification and an accurate online transverse momentum
measurement, in addition to single bunch-crossing identification with good time resolution.
The good mechanical precision of the chambers allows the track segments
to be reconstructed with a resolution better than $250~\mu \rm m$ \cite{MUON}.

A fundamental ingredient of the DT system is the calibration, which is
used as input to the local hit reconstruction, within the drift cells, and
thereby influences the precision of the track reconstruction.  This
paper describes in detail 
the DT calibration procedures, to obtain reliable calibration constants,
 and presents results obtained
 from the extensive commissioning run with cosmic ray events
performed in Autumn 2008, the Cosmic Run At Four Tesla (CRAFT)
\cite{Darin}, in preparation for LHC running.

The data sample used for
the calibration process and the trigger conditions are summarized in
Section 2.  The main characteristics of the DT calibration process are
described in Section~3. The process consists in the determination of
the inter-channel synchronization, described in Section 4, the analysis
of noisy channels, treated in Section 5, and the calculation of the
time pedestals and the drift velocity, described in Sections 6 and 7,
respectively.  The DT calibration workflow, including the monitoring
of the conditions, is performed within the CMS computing framework, as
described in Section 8.  Finally, a more refined analysis of
the drift velocity in the muon system, within the offline
reconstruction process, is presented in Section 9.

\section{Cosmic Ray Event Trigger and the Data Sample }

The long cosmic ray data taking in 2008  with and without magnetic
field allowed a detailed study of the DT  drift properties and an improved
understanding of the calibration constants. 
About 270 million events were collected with a $3.8~\rm T$ field
inside the solenoid magnet. In this configuration, the radial component of the magnetic field
in the DT chamber positions does not exceed $0.8~\rm T$.

The DT system was the primary trigger source for most of the collected
events.  The local trigger \cite{MUON} was
designed to operate with collisions taking place at the center 
of the CMS detector, and it is performed searching the 
$\phi$-matching of hits in each
chamber. This is achieved using dedicated hardware, which configures
the expected track paths from one chamber to another.

Due to the different origin, direction, and timing of the cosmic rays, as
compared to muons from proton-proton collisions, dedicated adjustments
were needed to properly configure the DT trigger for high efficiency
during CRAFT. 
%This required relaxing the extrapolation algorithm
%designed for LHC operation, to maximize the geometrical
%acceptance.  
This required relaxing the extrapolation algorithm 
with a particular configuration of the DTTF (DT Track Finder) 
as explained in more detail in Ref.~\cite{TriggerCraft}.
Therefore, during data-taking with cosmic rays, the L1 trigger was
generated by the coincidence in time of two segments in
two stations of the same sector, or adjacent sectors, 
and a rate of about $240~\rm Hz$ was provided to
the Global Muon Trigger.

About 20 million events, out of the 270 million collected during
CRAFT, were used for the calculation of the calibration
constants. They have been chosen from stable runs where most of the DT
system was operational. No quality cuts are, in principle, necessary to
perform the calibration. However, in order to have a clean sample of
muons, a transverse momentum cut of $7~\rm GeV$ was applied.

\section{The Calibration Process}

Charged particles crossing a DT cell produce ionization electrons in
the gas volume.  The determination of the relationship between the
arrival time of the ionization signal and its spatial deposition is
the primary goal of the calibration task, which leads to the
extraction of the drift times and drift velocities.

The arrival time of the ionization signal is measured using a high
performance Time to Digital Converter (TDC) \cite{TDC}. This is the
main building block of the read-out boards of the DT system. It is a
multi-hit device in which all hits within a programmable time window,
large enough to accommodate the cell maximum drift time, are assigned
to each Level 1 Accept trigger.  The drift time is directly obtained
from the time measured by the TDC, after subtracting a time pedestal
which contains contributions from the latency of the trigger and the
propagation time of the signal, within the detector and the data
acquisition chain.  The first goal of the calibration procedure is,
therefore, to determine the time pedestals, as described in Section 6.
The expected precision of the time pedestal calibration during 
the cosmic ray data-taking is limited by the arrival time distribution 
of cosmic rays which is flat within the clock cycle and it is of the 
order of ${25~\rm ns}\rm /{\sqrt{12}}$.

The other relevant quantity for the DT calibration process 
is the effective drift velocity. 
%The effective drift velocity 
It depends on many parameters, including the gas
purity and the electrostatic configuration of the cell, the
presence of a magnetic field within the chamber volume, and the
inclination of the track.
The parameters connected to the working conditions of the chambers 
%The working conditions of the chambers
are monitored continuously
\cite{MUON}: the high voltage supplies have a built-in monitor for
each channel; the gas is at room temperature  and its temperature
is measured on each preamplifier board inside the chamber; the gas
pressure is regulated and measured at the gas distribution rack on
each wheel, and is monitored by four further sensors placed at the
inlet and outlet of each chamber.  The adequacy of the flow sharing
from a single gas distribution rack to 50 chambers is monitored at the
inlet and outlet line of each individual chamber. A possible leakage in
the gas line can be sensed via the flow and/or the
pressure measurements.

Five small gas chambers, one for each wheel, are used to measure  
the drift velocity in a volume of very homogeneous electric field, 
located in the accessible gas room adjacent to the cavern, outside of
the CMS magnetic field.
Each of these chambers, called Velocity Drift Chambers (VDC)~\cite{diplom-altenhoefer,diplom-frangenheim}, is
able to selectively measure the gas being sent to, and returned from,
each individual chamber of the wheel
thus providing rapid feedback on any changes due to the gas mixture
or contamination.
During the CRAFT data-taking period only one such chamber was used.

No noticeable variation of  the parameters described above is expected among
different regions of the spectrometer. However, the magnetic
field and the track impact angle may vary substantially from chamber
to chamber, as they occupy different positions in the return yoke.

Two methods for calculating the electron drift velocity in a DT cell
are presented here. They both assume a constant drift velocity in a
given chamber. The first method, discussed in Section~7, is based on
the measurement of the effective drift velocity using the mean-time
technique, which computes the velocity value at the super-layer
granularity level. The second method, discussed in Section 9,  relies
on the muon track fit, which determines track-by-track the time of
passage of the muon and the drift velocity as additional free
parameters of the fit, together with  the track position and
inclination angle.
The assumption of a constant drift velocity is considered a good
approximation because the magnetic field in the chamber volume is
usually low and approximately homogeneous.
A  third method based on a parameterization of the drift velocity as a
function of the drift time, the magnetic field, and the muon
trajectory is discussed in Ref.~\cite{vDriftPar}.

The detailed drift velocity analysis, described in Section 9, also reveals
non-linear effects in the innermost stations (MB1 chambers) of the
barrel external wheels (Wheel +2 and Wheel -2), where the strongest
radial magnetic field component, of about $0.8~\rm T$, is present.

The DT calibration process also depends on the different signal path
lengths to the read-out electronics (called inter-channel
synchronization time) and on the list of noisy channels, as will be
described in the following sections.

\section{The Inter-Channel Synchronization}

The inter-channel synchronization is calculated for each read-out channel of
each chamber, in order to correct for the different signal path
lengths of trigger and read-out electronics.  This is a fixed offset,
since it only depends on cable/fiber lengths, and it does not need to
be re-calibrated very often.  Nevertheless, it is useful to frequently
redo its calibration, to monitor the correct behavior of the front-end
electronics.  The inter-channel synchronization is determined by
test-pulse calibration runs.  The design of the data acquisition
system allows such runs to be taken during the normal physics
data-taking, by exploiting the collision-free interval of the LHC beam
structure, called ``abort gap''.

During special calibration runs, a test-pulse is simultaneously
injected in four channels of a front-end board, each one from a
different layer of a super-layer, simulating a muon crossing the
super-layer. To perform the scanning of the entire DT system in only
16 cycles, the same test-pulse signal is also distributed to other
four-channel groups, 16 channels apart.

The so-called $t_0$ calibration consists in determining, for each DT
channel, the mean time and the standard deviation of the test-pulse.
In the calibration procedure, the events are split in two samples: the
first is used to compute the average value, within a full chamber, of
the signal propagation time from the test-pulse injector to the
read-out electronics; the second is used to calculate, for each
individual channel, the difference between the time of its test-pulse
signal and the average value of the chamber.

\begin{figure}[htbp]
 \begin{center}
   \resizebox{10cm}{!}{\includegraphics{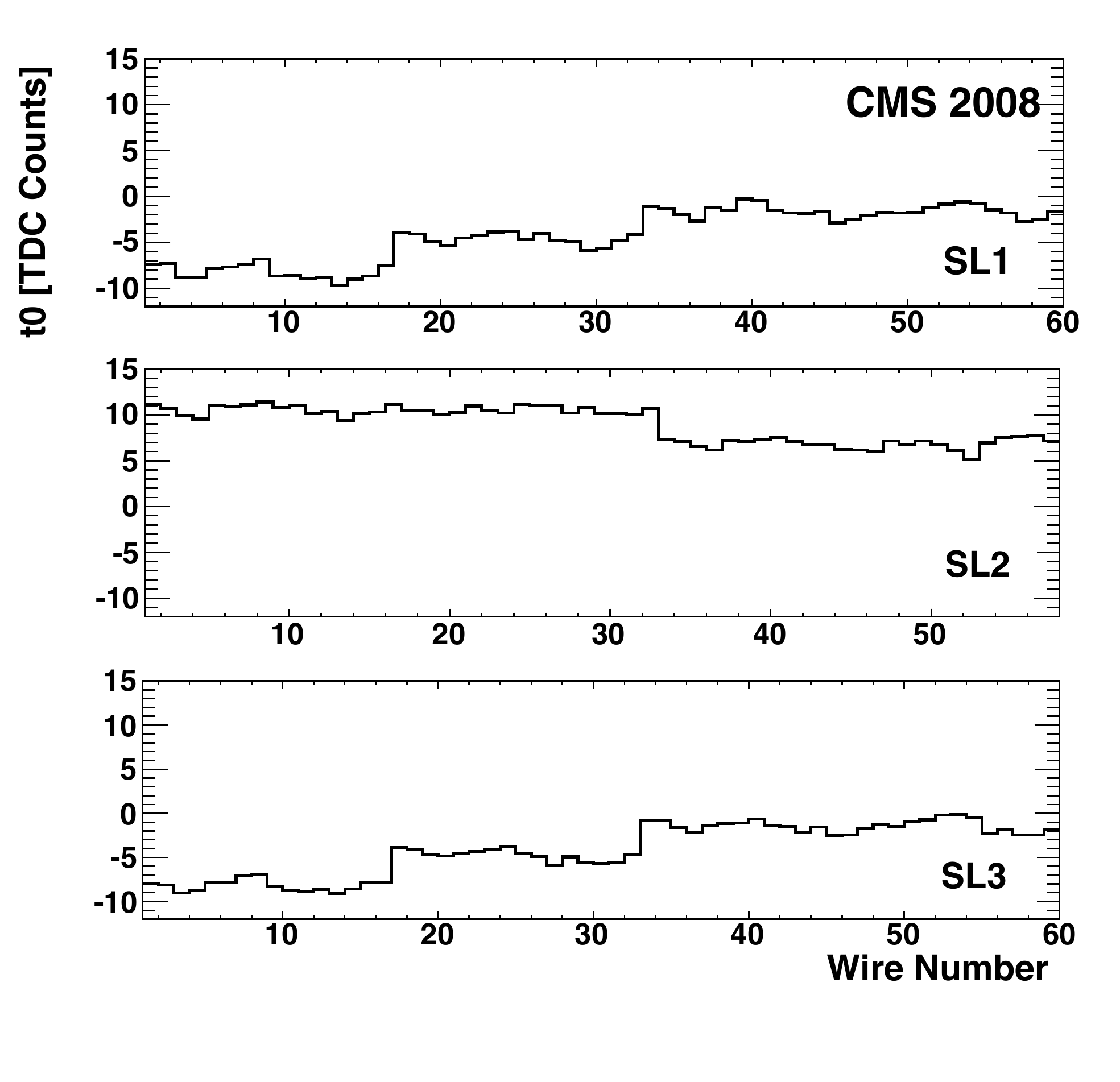}}
   \caption{Inter-channel synchronization constants calculated from a
     test-pulse run.  The results are shown for three representative
     layers, belonging to each of the three super-layers of
     chamber MB3 in Sector 9.  The step-function shape reflects the
     grouping of channels among different front-end boards.}
  \label{fig:t0TestPulse}
 \end{center}
\end{figure}

Figure~\ref{fig:t0TestPulse} shows an example of a distribution of
$t_0$ constants for representative layers of the three super-layers of
a chamber, as a function of the channel number.  The other layers show
very similar $t_0$ values.  The $t_0$ synchronization correction is
always below $10~\rm ns$ (1 TDC count corresponds to $0.78~\rm ns$).
The standard deviation is about $1~\rm ns$, for all channels. This is
compatible with the  precision of the electronic chain.
 These corrections correspond to the distance between the front-end boards,
located inside the chamber volume, and the read-out boards.  Wires
connected to a given front-end board belong to cells adjacent to each
other, in the super-layer, and have approximately the same distance up
to the read-out boards. This leads to the step-function shape seen in
Fig.~\ref{fig:t0TestPulse}, more pronounced in the super-layers 1 and
3.

\section{Noise Analysis in the DT Chambers}

On the basis of systematic studies performed during several
commissioning phases of the DT detector, a cell is defined as noisy if
its hit rate at operating voltage, counting signals higher than a common discriminator
threshold of $30~\rm mV$, is higher than $500~\rm Hz$.

\begin{figure}[htbp]
  \begin{center}
   \resizebox{14cm}{!}{\includegraphics{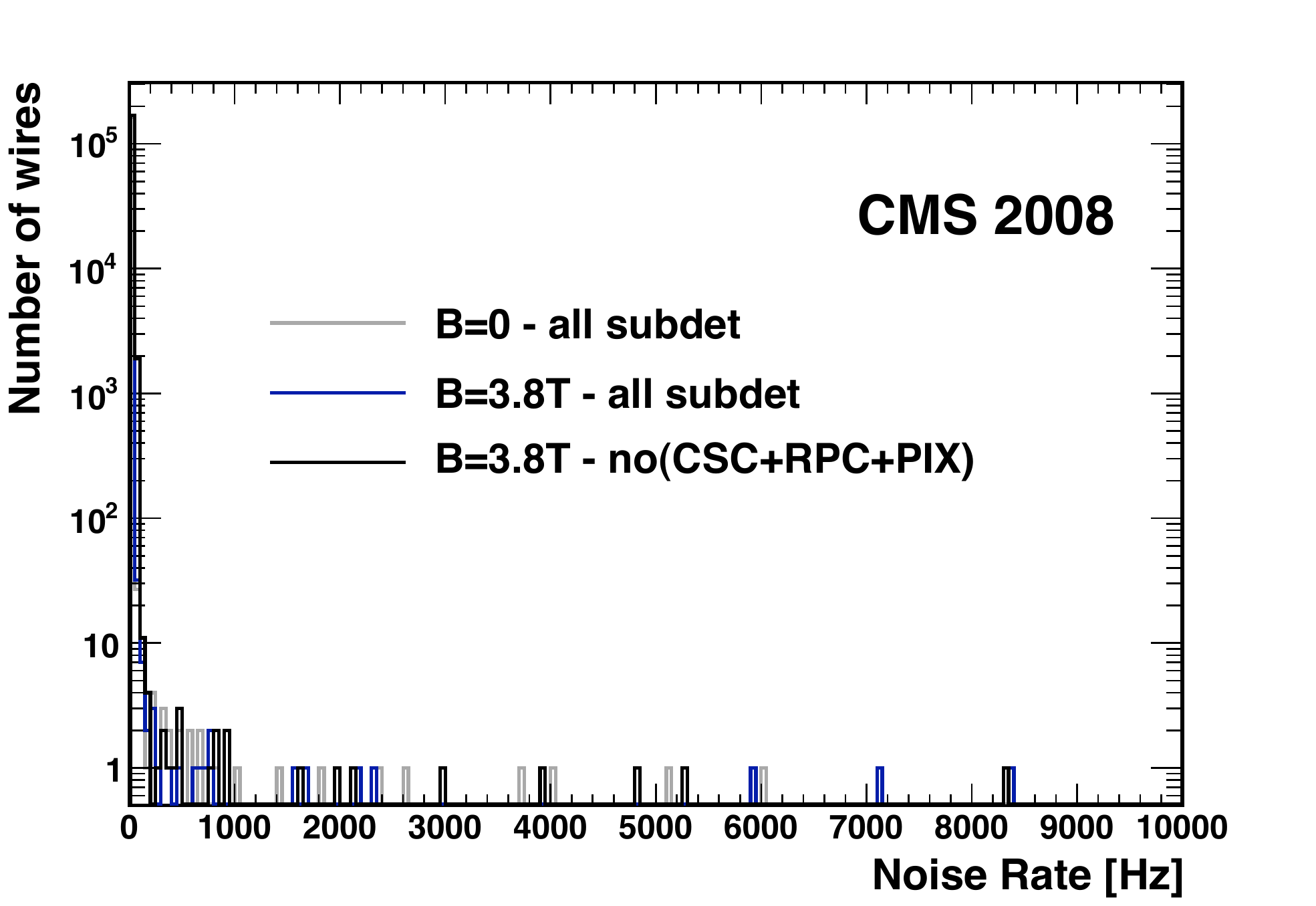}}
   \caption{Distribution of the cell noise rate for different
     data-taking conditions: with and without magnetic field; with and
     without Cathode Strip Chambers (CSC), Resistive Plate Chambers
     (RPC), and Pixels (PIX).}
   \label{fig:noiseRate}
  \end{center}
\end{figure}

The number and geometrical distribution of noisy DT channels have been
studied, in particular, during a commissioning period without magnetic
field and having the detector wheels separated from each other.
In this section we describe the results of the noise analysis
performed using the cosmic ray events collected during CRAFT.  With
respect to runs using random triggers, the noise analysis based on
normal data taking runs has the advantage of reflecting the detector
operation in more realistic conditions. 

The first aim of the noise studies is to check the stability of the
number of noisy cells in different conditions of the CMS detector.
For all runs analyzed, the number of noisy cells is around $0.01~\%$
of all DT channels.  The rate of noise hits per cell is shown in
Fig.~\ref{fig:noiseRate}, for a number of representative runs, with
and without magnetic field, and 
with different sub-detectors included in the acquisition. 
 The number of cells with a hit rate higher than
$500~\rm Hz$ is very small.  Detailed information on the
noise rate observed for representative runs, with different
configurations of CMS detectors included in the data acquisition, is
shown in Table ~\ref{table:TabNoise1}.  

\begin{table}[htbp]
 \centering
  
\caption{Number of Noisy Cells in each chamber type (MB1, MB2, MB3,
  MB4) and average noise rate for some representative runs of
  different data-taking conditions.  The results from the CRUZET
  (Cosmic RUn at ZEro Tesla) commissioning period are also shown, for
  comparison.}
 
  \begin{tabular}{|c|c|c|c|c|c|c|c|}
     \hline
     \multicolumn{1}{|c|}{\textbf{Data}} &
     \multicolumn{1}{c|}{\textbf{B Field}} &
      \multicolumn{1}{c|}{\textbf{Excluded }} & 
      \multicolumn{1}{c|}{\textbf{Number}} &
      \multicolumn{1}{c|}{\textbf{Number}} &
      \multicolumn{1}{c|}{\textbf{Number}} &
      \multicolumn{1}{c|}{\textbf{Number}} &
        \multicolumn{1}{c|}{\textbf{Mean }}        \\

     \multicolumn{1}{|c|}{\textbf{Period}} &
     \multicolumn{1}{c|}{\textbf{ [T] }} &
      \multicolumn{1}{c|}{\textbf{Sub-Det.}} & 
      \multicolumn{1}{c|}{\textbf{Noisy}} &
      \multicolumn{1}{c|}{\textbf{Noisy}} &
      \multicolumn{1}{c|}{\textbf{Noisy}} &
      \multicolumn{1}{c|}{\textbf{Noisy}} &
        \multicolumn{1}{c|}{\textbf{Noise Rate}}        \\
 
     \multicolumn{1}{|c|}{\textbf{}} &
     \multicolumn{1}{c|}{\textbf{ }} &
      \multicolumn{1}{c|}{\textbf{}} & 
      \multicolumn{1}{c|}{\textbf{Cells}} &
      \multicolumn{1}{c|}{\textbf{Cells}} &
      \multicolumn{1}{c|}{\textbf{Cells}} &
      \multicolumn{1}{c|}{\textbf{Cells}} &
        \multicolumn{1}{c|}{\textbf{}}        \\

     \multicolumn{1}{|c|}{\textbf{ }} &
     \multicolumn{1}{c|}{\textbf{ }} &
      \multicolumn{1}{c|}{\textbf{ }} & 
      \multicolumn{1}{c|}{\textbf{in MB1}} &
      \multicolumn{1}{c|}{\textbf{in MB2}} &
      \multicolumn{1}{c|}{\textbf{in MB3}} &
      \multicolumn{1}{c|}{\textbf{in MB4}} &
        \multicolumn{1}{c|}{\textbf{[Hz] }}        \\

    \hline
     CRUZET & 0 & CSC, PIX & 12 & 2 & 2 & 0 & 3.96    \\
     CRAFT  & 0 &  & 0  & 13 & 3 & 2 & 3.80   \\
     CRAFT  & 3.8 &  & 13  & 3 & 1 & 2 & 4.15   \\
     CRAFT  & 3.8 & CSC & 12  & 7 & 0 & 8 & 4.23   \\
     CRAFT  & 3.8 & CSC,& 17 & 5 & 0 & 0 & 4.50   \\
            &  &  PIX, RPC&  &  &  &  &    \\
 \hline
   \end{tabular}
  
 \label{table:TabNoise1}
\end{table} 

An average noise rate of $\sim 4~\rm Hz$ is observed in the DT system,
essentially insensitive to the magnetic field and to the status of
nearby sub-detectors.
In addition, it has been observed that around $50~\%$ of the noisy
cells remain noisy for long data-taking periods.

\begin{figure}[htbp]
  \begin{center}
   \resizebox{14cm}{!}{\includegraphics{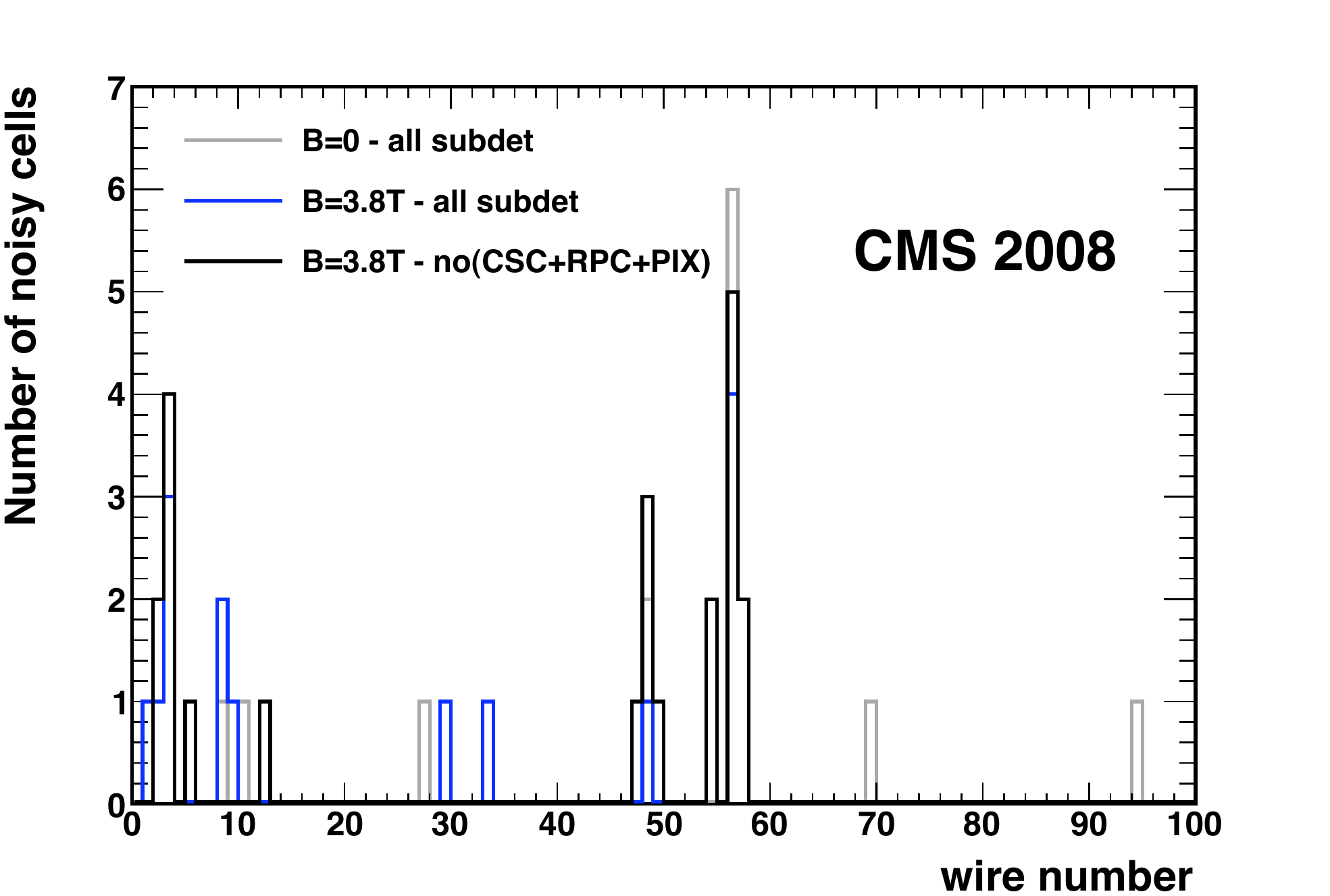}}
   \caption{Number of noisy channels as a function of the wire number,
     for the data-taking periods mentioned in
     Table~\ref{table:TabNoise1}. Each distribution corresponds to a
     different run and shows the noisy cells observed for chamber
     types MB1, MB2, MB3 and MB4.}
   \label{fig:noisyCells}
  \end{center}
\end{figure}

Studies have also been made concerning the position dependence of the
noisy cells, within wheels and chambers.  As seen on
Table~\ref{table:TabNoise1}, most of the noisy channels are located in
the innermost chambers (MB1), where the internal cabling is more
complex, because of the reduced space.  In Fig.~\ref{fig:noisyCells},
the noisy cell distribution is shown as a function of the wire number.
The noisy cells appear concentrated in the regions of the super-layer
close to the wire boundaries, which are different depending on the
number of wires present in each chamber type (about 50 for MB1, 60 for
MB2, 70 for MB3, and 90 for MB4).  The peaks also reflect the position
of the connectors distributing the HV inside the super-layer, which
generate some electronic noise in their proximity.

The observed fraction of noisy cells ($0.01~\%$) and the average noise
rate ($\sim 4~\rm Hz$) in the full DT system are too low to affect the
digitization efficiency or the trigger rate. It is important, however,
to exclude the noisy cells in the calibration process described in the
following sections.

\section{The Time Pedestal Calibration}

\subsection{Computation of the Calibration Constants}

The time pedestal calibration is the process which allows the
extraction of the drift time from the TDC measurement.  For an ideal
drift cell, the time distribution coming from the TDC ($t_\textrm{TDC}$)
would coincide with the distribution of drift time ($t_\textrm{drift}$), and
would have a box shape starting from a null drift time, for tracks
passing near the anode, up to about $380~\rm ns$, for tracks passing
near the cathode.

Experimentally, some non-linear effects related to the electric field
distribution inside the drift cell have to be considered in the
response of these cells; they are enhanced by the track
inclination and by the presence of the magnetic field.  In addition,
different time delays, related to trigger latency, and different cable
lengths of the read-out electronics, also contribute to the TDC
measurements.  The time measured by the TDC, $t_\textrm{TDC}$, can be
expressed as

\begin{equation}
t_\textrm{TDC} = t_0 + t_\textrm{TOF} + t_\textrm{prop} +
t_\textrm{L1} + t_\textrm{drift} \quad ,
\label{eq1}
\end{equation}

where
\begin{itemize}
    \item $t_0$ is the inter-channel synchronization used to equalize the
      response of all the channels at the level of each chamber, as
      described in Section 4;
    \item $t_\textrm{TOF}$ is the Time-Of-Flight (TOF) of the muon, from the
    interaction point to the cell, in the case of collision events. In
    the case of cosmic events, this quantity cannot be defined because
    the time pedestal is an average of the arrival time
    of cosmic muons relative to the clock cycle;
  \item $t_\textrm{prop}$ is the propagation time of the signal along the
    anode wire;
    \item $t_\textrm{L1}$ is the latency of the Level-1 trigger;
    \item $t_\textrm{drift}$ is the drift time of the electrons from the
    ionization cluster to the anode wire within the cell.
\end{itemize}

  \begin{figure}[htbp]
  \begin{center}
   \resizebox{12cm}{!}{\includegraphics{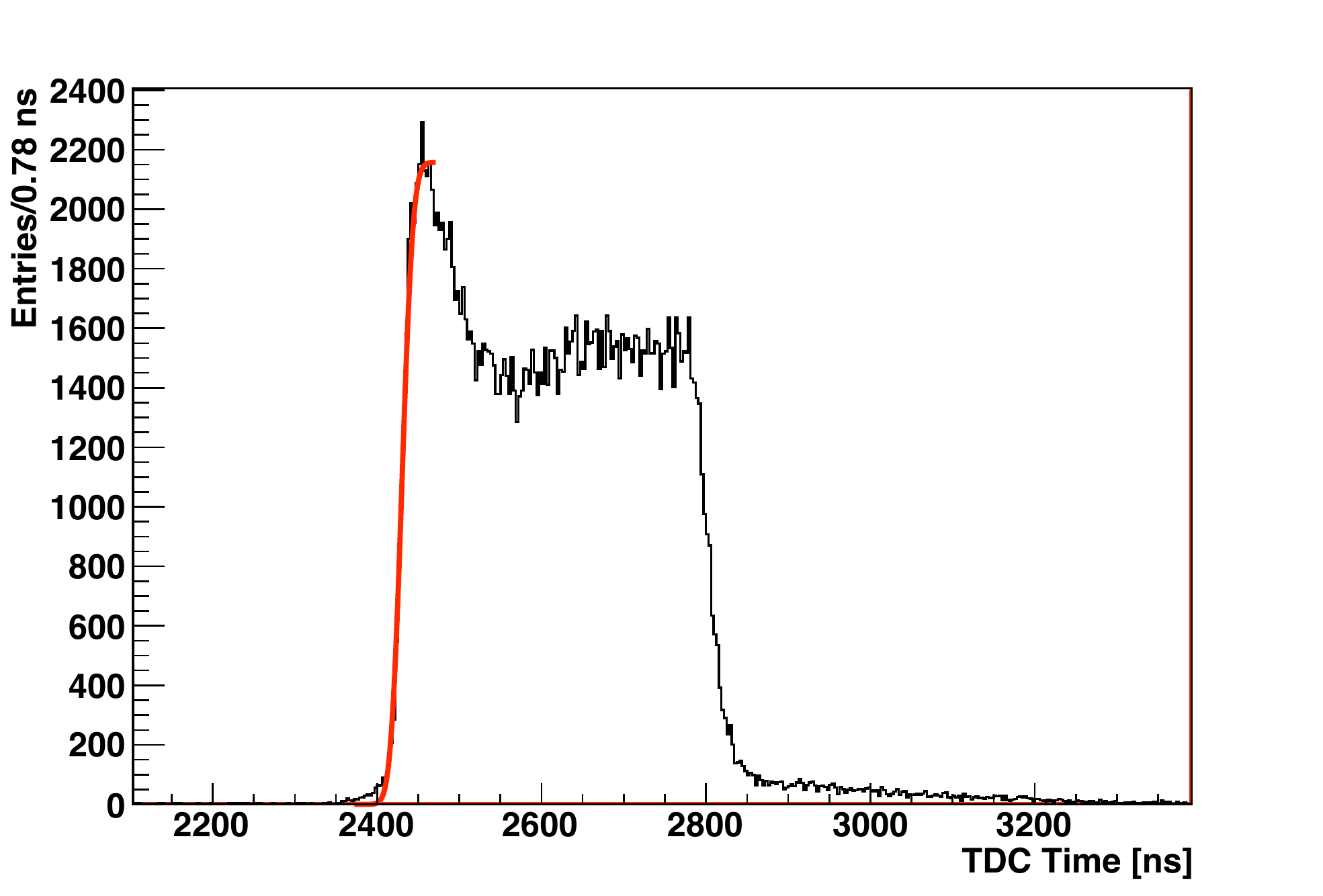}}
   \caption{Distribution of the signal arrival times, recorded by the
     TDC, for all the cells of a single super-layer in a chamber,
     after the cell-to-cell equalization based on the test-pulse
     calibration. The continuous line indicates the fit of the
     \emph{Time Box} rising edge to the integral of a Gaussian
     function.}
   \label{fig:TimeBox}
  \end{center}
\end{figure}

The main goal of the calibration is the calculation of the 
%three terms that contribute to the 
time pedestal, $t_\textrm{trig}$, which is dominated by the time delay
caused by the L1 trigger latency:

\begin{equation}
t_\textrm{trig} = t_\textrm{TOF} + t_\textrm{prop} + t_\textrm{L1}
\quad .
\label{eq1.1}
\end{equation}

The value of $t_\textrm{trig}$ is extracted for each super-layer 
directly from the
$t_\textrm{TDC}$ distribution, referred to as \emph{Time Box}, after
subtracting the noisy channels and correcting for the inter-channel
synchronization. Figure~\ref{fig:TimeBox} shows a \emph{Time Box}
measured during a CRAFT run, for one super-layer.

The value of $t_\textrm{trig}$ is the turn-on point of the \emph{Time
  Box} distribution. It is computed by fitting the rising edge of the
distribution to the integral of a Gaussian function, as illustrated by
the continuous line in Fig.~\ref{fig:TimeBox}. The procedure is
applied at the super-layer level and is described in more detail in
Ref.~\cite{CalProc}.

\begin{figure}[htbp]
\begin{center}
\begin{tabular}{c}
\resizebox{12cm}{!}{\includegraphics{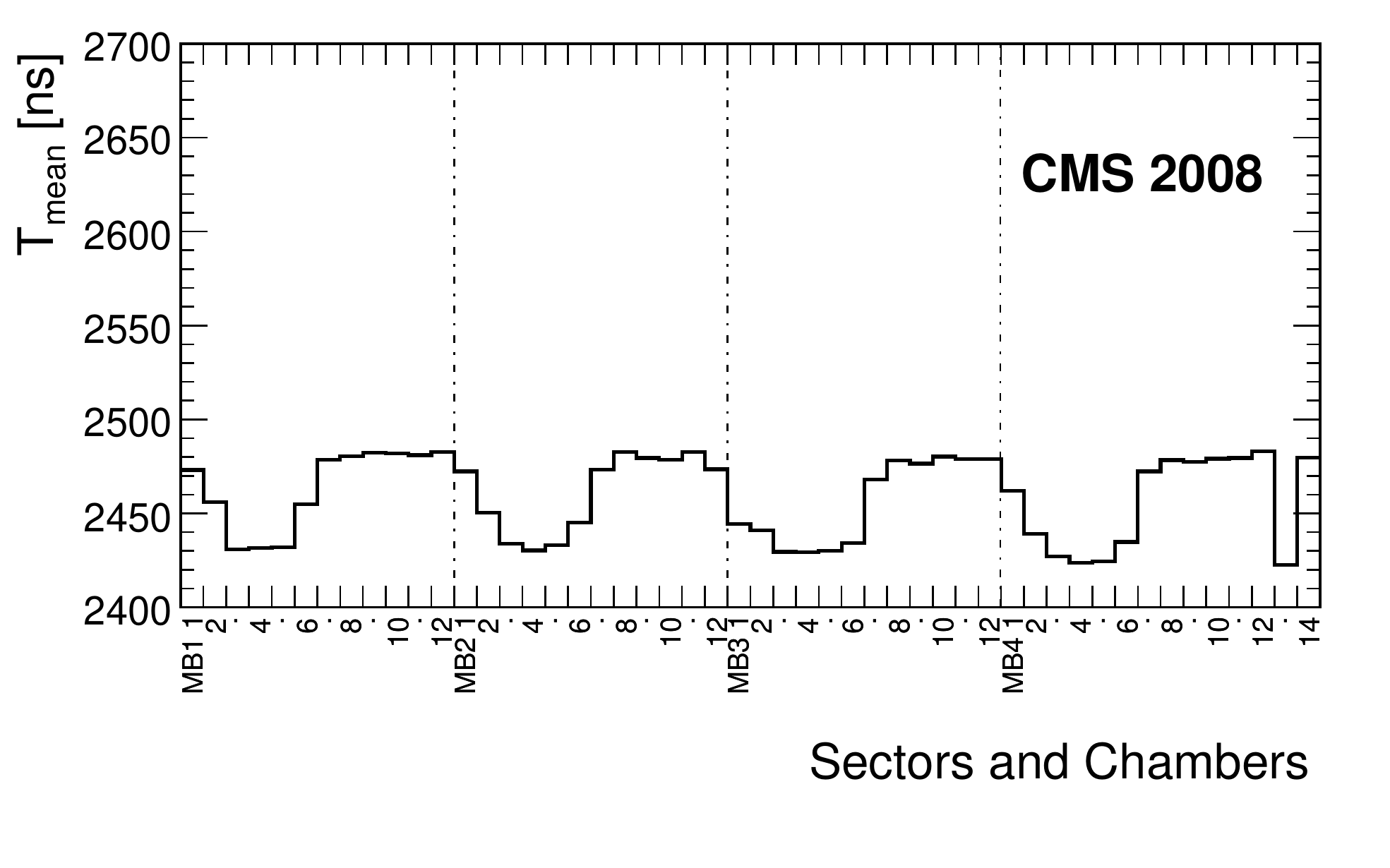}} \\
\resizebox{12cm}{!}{\includegraphics{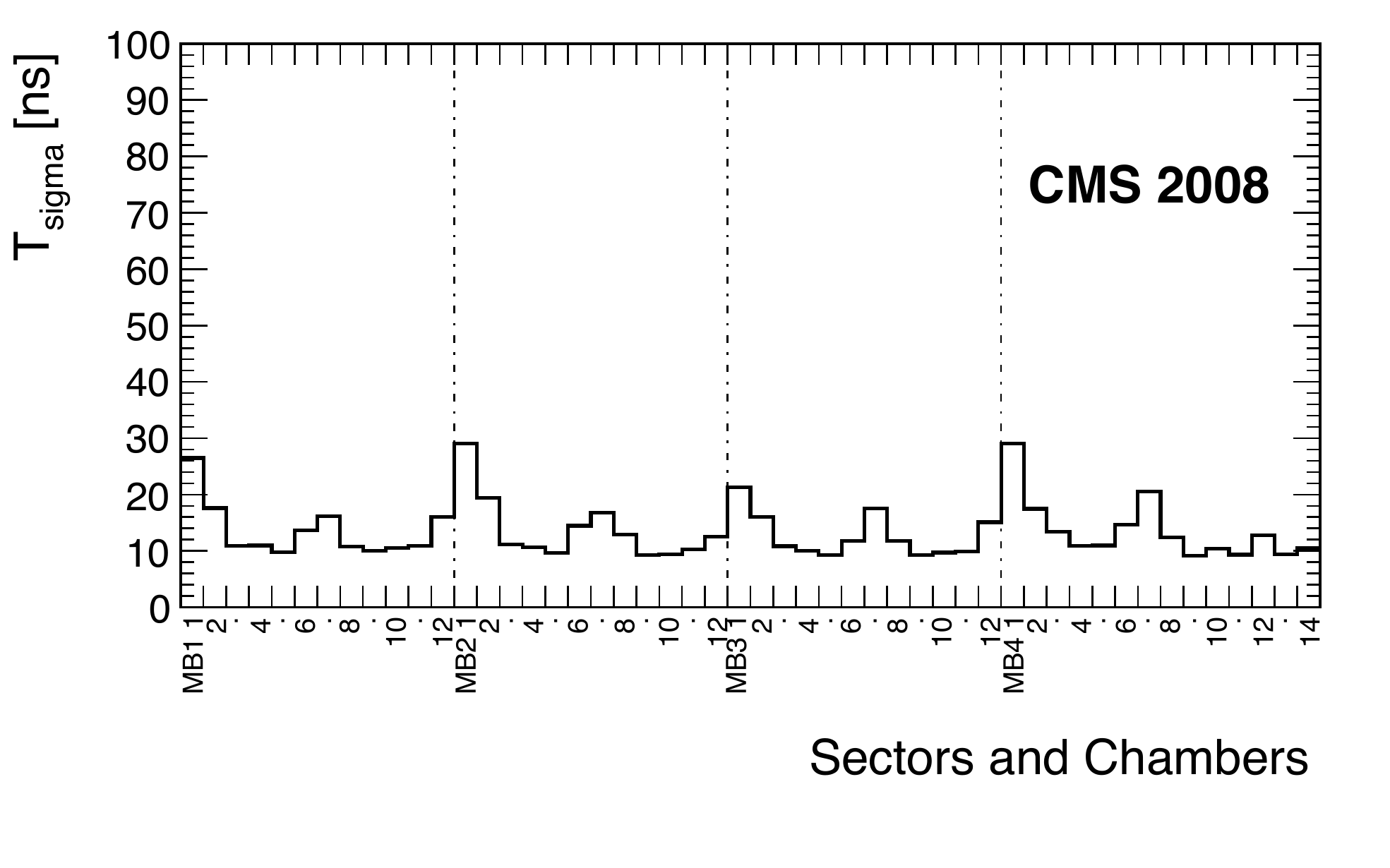}} \\
\end{tabular}
\caption{Mean (top) and standard deviation (bottom) of the fitted
  inflexion point of the \emph{Time Box} rising edge, for the
  innermost $r\phi$ super-layer for a representative wheel. The
  triggering sectors (3, 4, 5 and 9, 10, 11) are synchronized among
  each other.  The sectors with vertical chambers (sectors 1 and 7)
  detect much less cosmic ray muons, leading to a poorly defined
  rising edge and a less accurate calibration.}
\label{fig:Fig89a}
\end{center}
\end{figure}

The main quantities calculated by the fit are the inflexion point of
the rising edge, $T_\textrm{mean}$, and its standard deviation,
$T_\textrm{sigma}$, which represents the resolution of the
measurement.  Figure~\ref{fig:Fig89a} shows the distributions of
$T_\textrm{mean}$ and $T_\textrm{sigma}$ measured, in a CRAFT run with
$B=3.8~\rm T$, for the innermost $r\phi$ super-layer of a
representative wheel, as a function of chamber type and sector.
Similar results were obtained for the other wheels.  Approximately
constant values are observed for chambers of the same type and for all
the wheels.  The periodic structure seen in the $T_\textrm{mean}$
distribution, Fig.~\ref{fig:Fig89a} (top), reflects the time-of-flight
of the cosmic muon from the upper sector to the lower sector.
Indeed, the events contributing to the calculation of
$T_\textrm{mean}$ and $T_\textrm{sigma}$ can be triggered by the upper
or lower sectors. The events triggered by the top (bottom) sectors may also be
detected by other non-triggering sectors, having a less precise time
pedestal and, consequently, leading to a less precise determination of
these quantities.  Different runs during the entire CRAFT period have
been analyzed, and a stable performance of the whole DT system has
been observed. As expected, no dependence on the magnetic field
strength was observed.

The time resolution distribution, Fig.~\ref{fig:Fig89a} (bottom),
indicates the precision which the calibration procedure can reach with
cosmic rays. 
A standard deviation of $\sim 10~\rm ns$ is observed for all
super-layers in all wheels, except in the vertical sectors, where the
number of events is limited and the muon crossing angles are large.
The time resolution precision is limited mainly by the random arrival 
time of cosmic muons relative to the clock cycle. 
Furthermore, the resolution in Sector 1 is systematically worse than
in Sector 7 because the trigger cables that distribute the Level 1
accept signal are longer and, therefore, generate larger skews in the
signal transmission.

After the determination of $T_\textrm{mean}$ and $T_\textrm{sigma}$,
the time pedestal, $t_\textrm{trig}$, is estimated as

\begin{equation}
t_\textrm{trig} = T_\textrm{mean} - k \cdot T_\textrm{sigma} \quad.
\label{eq2}
\end{equation}

The $k$ factor is evaluated by minimizing the position residuals,
using the local reconstruction of track segments within chambers.
After a few iterations, a $k$ factor of 0.7 was computed for the CRAFT
data and was applied to all super-layers.  The position residuals were
then recalculated and a final correction to the time pedestals was
computed dividing the remaining offsets observed in the residual
distributions by a constant drift velocity ($54.3~ \mu \rm m/ns$).
The final $t_\textrm{trig}$ constants were stored in a database, as
described in Section 8.

\subsection{Validation of the Calibration Constants}

Once the $t_\textrm{trig}$ constants are computed, the calibration
process proceeds with the validation step, which consists in studying
the effect of these constants on the reconstruction algorithm.  The
analyzed quantities are the residuals computed, layer by layer, as the 
distance between the hit and the intersection of the 3D segment
with the layer plane.
A complete description of the local reconstruction procedure is given
in Ref.~\cite{DTReco}. 
  To correct for the propagation time along the wire, 
the reconstruction of the segment is done in a multistep procedure. 
 First the reconstruction is performed in the $r\phi$ and $z$
projections independently. Once two projections are paired and the
position of the segment inside the chamber is approximately known, the
 drift time is corrected for the propagation time along the wire and
for the TOF within the super-layer, and the 3D position is updated.  

\begin{figure}[htbp]
\begin{center}
\begin{tabular}{c}
\resizebox{12cm}{!}{\includegraphics{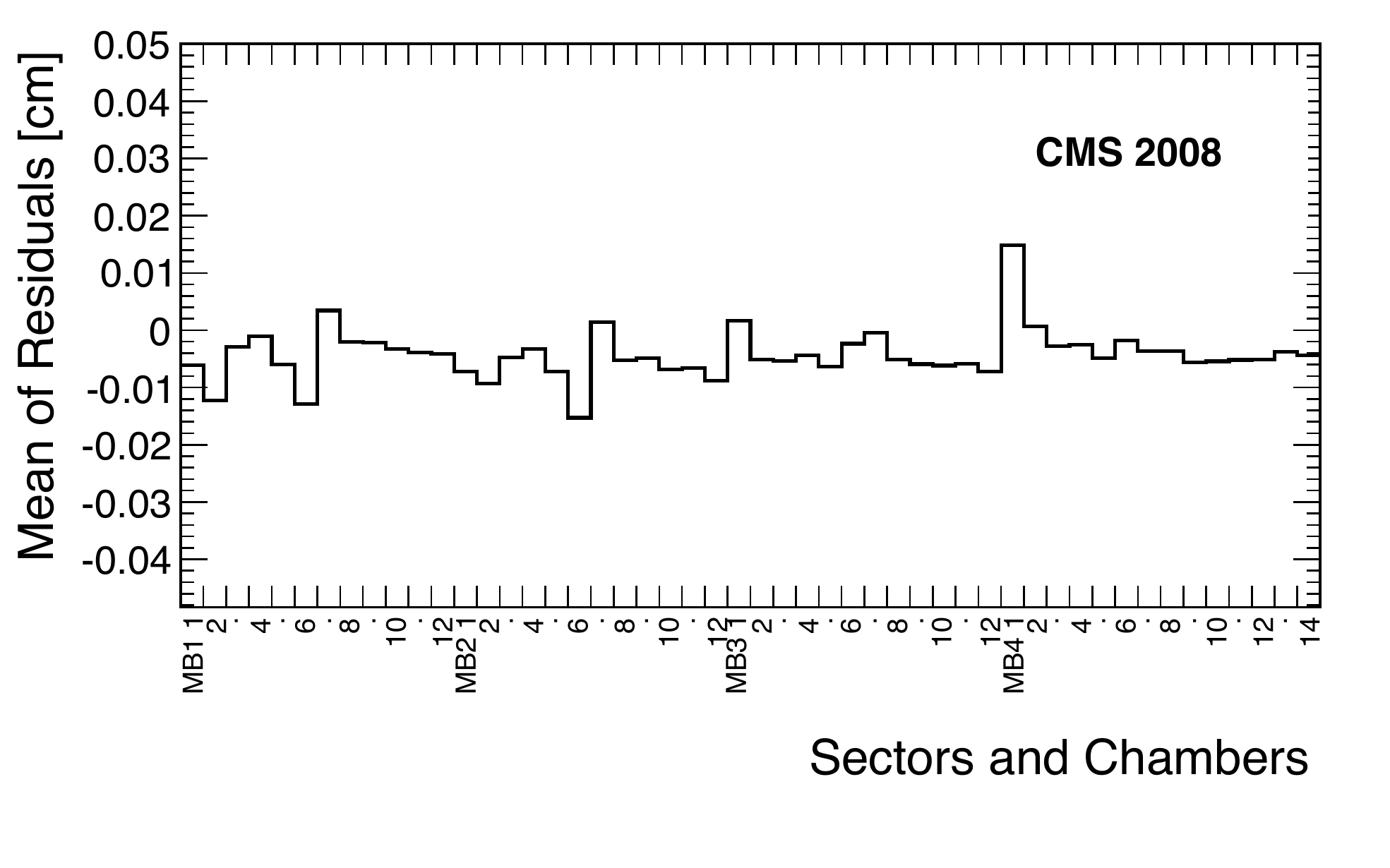}} \\
\resizebox{12cm}{!}{\includegraphics{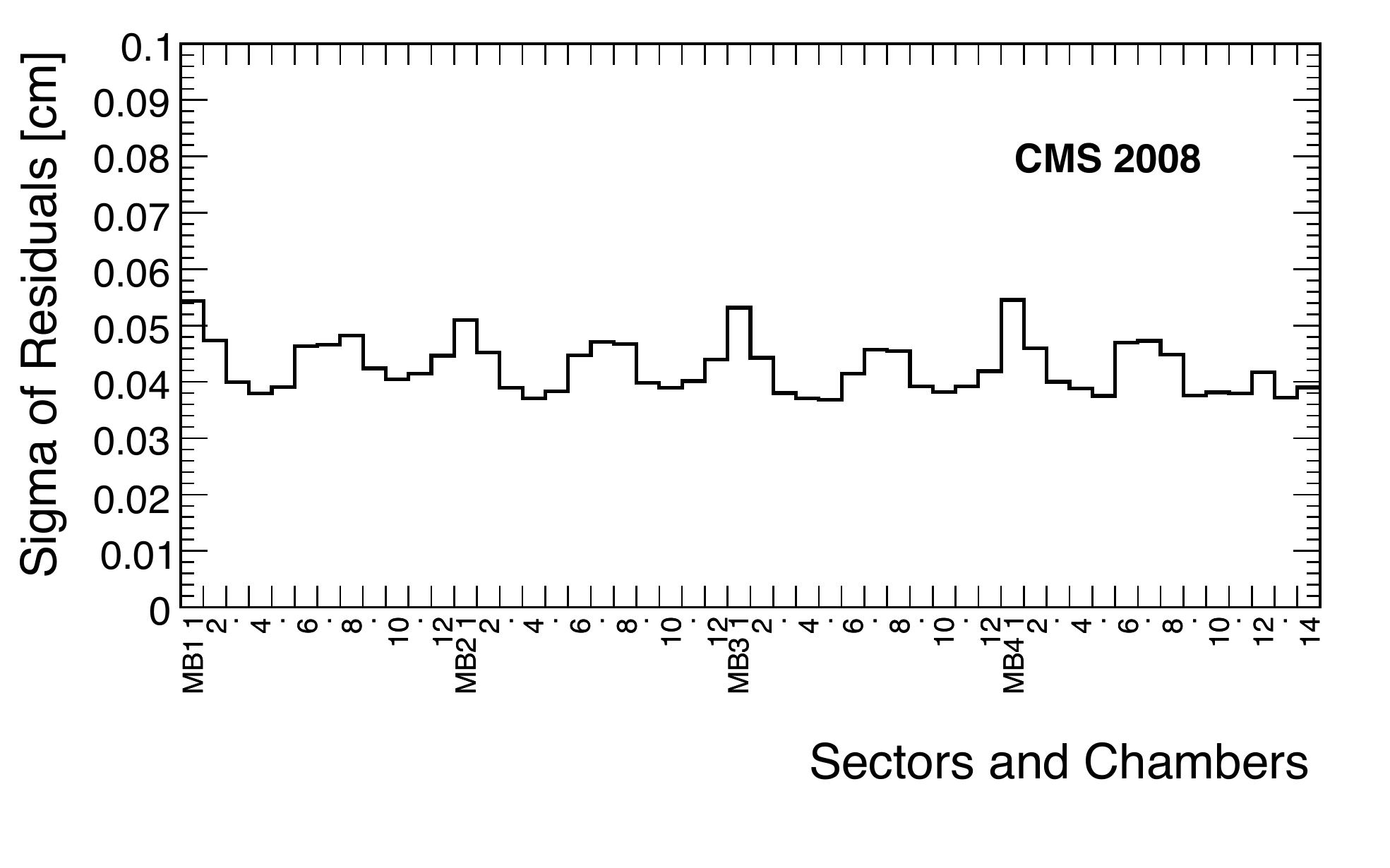}} \\
\end{tabular}
\caption{Mean (top) and width (bottom) Gaussian parameters, as fitted
  to the distributions of the residuals between the reconstructed hits
  and the reconstructed local segments. The results are shown for the
  $r\phi$ super-layers for a representative wheel, after correcting
  the offset with respect to the origin of the residual distribution.
  Other wheels show similar results.}
 \label{fig:Fig1213}
\end{center}
\end{figure}

The mean values of the residual distributions 
calculated for the innermost $r\phi$ super-layer 
for a representative wheel are shown in Fig.~\ref{fig:Fig1213} (top). 
A systematic offset with respect to the origin is observed 
compatible with the systematic delay between the arrival time of the
cosmic muon events and the clock cycle. 
%The distribution of the mean of the residuals, shown in
%Fig.~\ref{fig:Fig1213} (top), has an offset of the peak with respect
%to the origin compatible with the precision that can be reached with
%cosmic muon events, limited by their random arrival time relative to
%the clock cycle.  
The standard deviations of the fit to the residuals,
shown in Fig.~\ref{fig:Fig1213} (bottom) and in the range
400--600~$\mu \rm m$, represent the spatial resolution obtained with
the calibration process, a factor of two worse than the nominal
resolution of about $250~ \mu \rm m$~\cite{MUON}.  The difference is
caused by the spread of the muon arrival times inside the $25~\rm ns$
time window associated with the L1 trigger. This dilution will not
occur with LHC collision data.

\section{The Drift Velocity Calibration}

The aim of the drift velocity calibration is to find the best
effective drift velocity in each region of the DT system. In order to
be consistent 
with the $t_\textrm{trig}$ calculation, described in Section~6, the
drift-velocity calibration is computed with a super-layer granularity.

\begin{figure}[htbp]
  \begin{center}
   \resizebox{10cm}{!}{\includegraphics{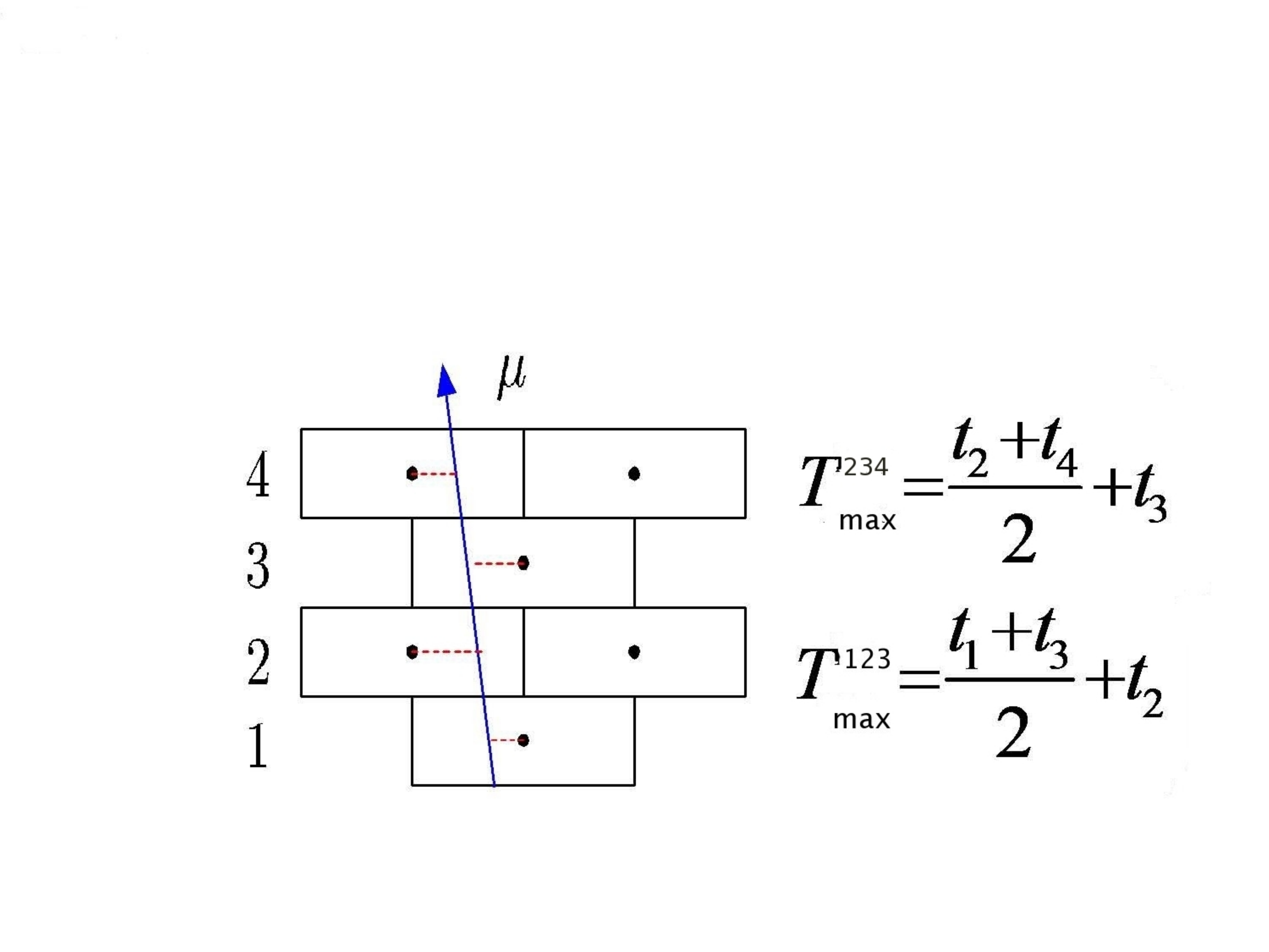}}
   \caption{Schematic view of a super-layer section, showing the
     pattern of semi-cells crossed by a track. The quantities $t_i$ in
     the equations represent the arrival time of the electrons within
     a drift cell.}
   \label{fig:MeanTime}
  \end{center}
\end{figure}

The calibration algorithm is based on the mean-time technique
described in detail in Ref.~\cite{CalProc}.  In this method, the
maximum drift time in a cell, $T_\textrm{max}$, is calculated
considering nearby cells in three adjacent layers and using a linear
approximation to determine the average drift velocity.  As an example,
Fig.~\ref{fig:MeanTime} shows the simplest pattern of a muon crossing
a semi-column of cells, together with the equations used to calculate
$T_\textrm{max}$.  In general, $T_\textrm{max}$ depends on the track
inclination and on the pattern of cells crossed by the track.  Taking
into account these dependencies, a spread of about $28~\rm ns$ has
been observed in the calculation of $T_\textrm{max}$ from CRAFT data.

The effective drift velocity can be estimated assuming a linear
space-time relationship,
\begin{equation}
v_\textrm{drift}^\textrm{eff} = \frac{L_\textrm{semi-cell}}{<T_\textrm{max}>} \quad ,
\label{eq3}
\end{equation}
where $L_\textrm{semi-cell} = 2.1~\rm cm$ is half the width of a
drift cell.

\begin{figure}[htbp] 
\begin{center}
\begin{tabular}{c}
\resizebox{12cm}{!}{\includegraphics{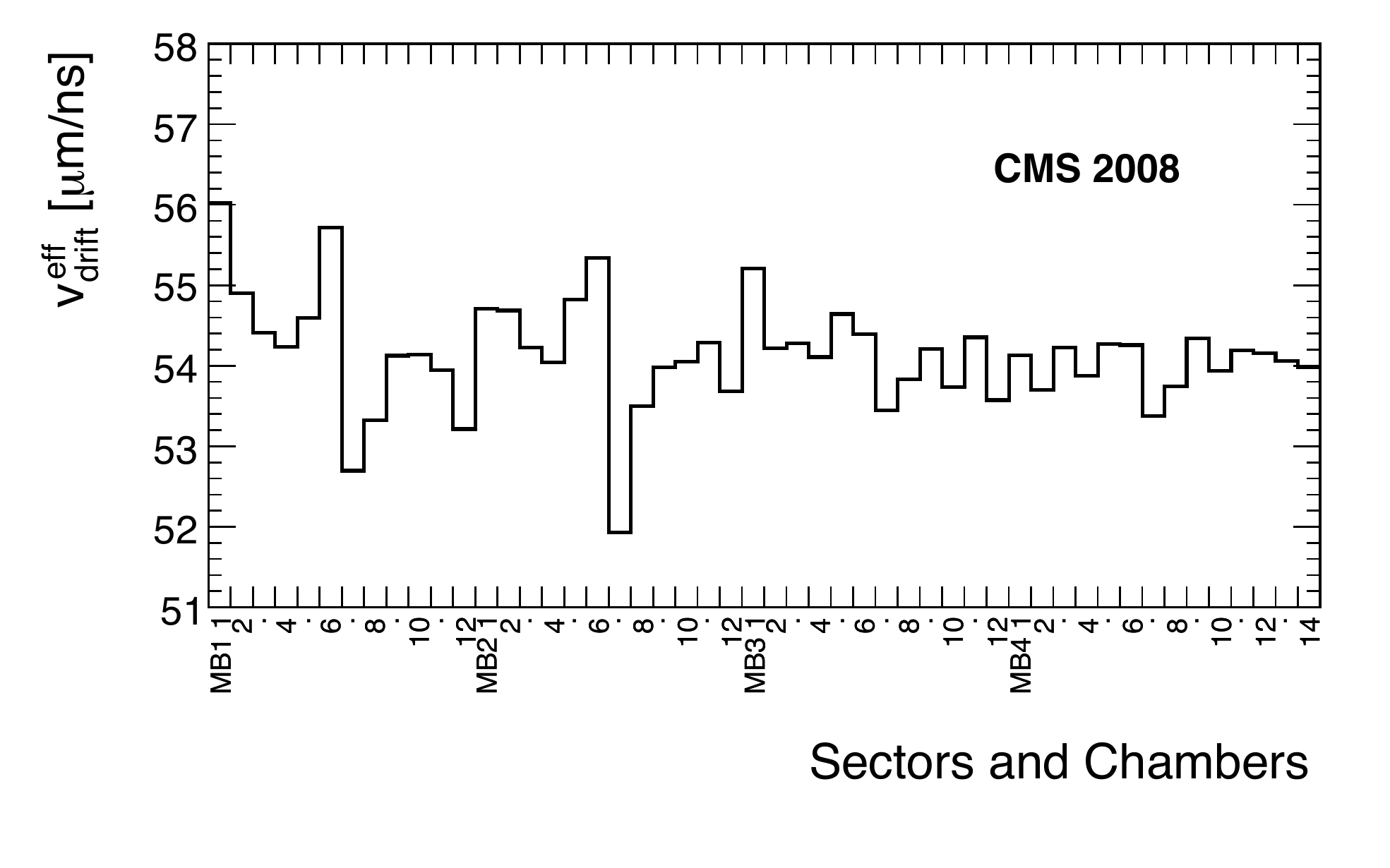}} \\ 
\resizebox{12cm}{!}{\includegraphics{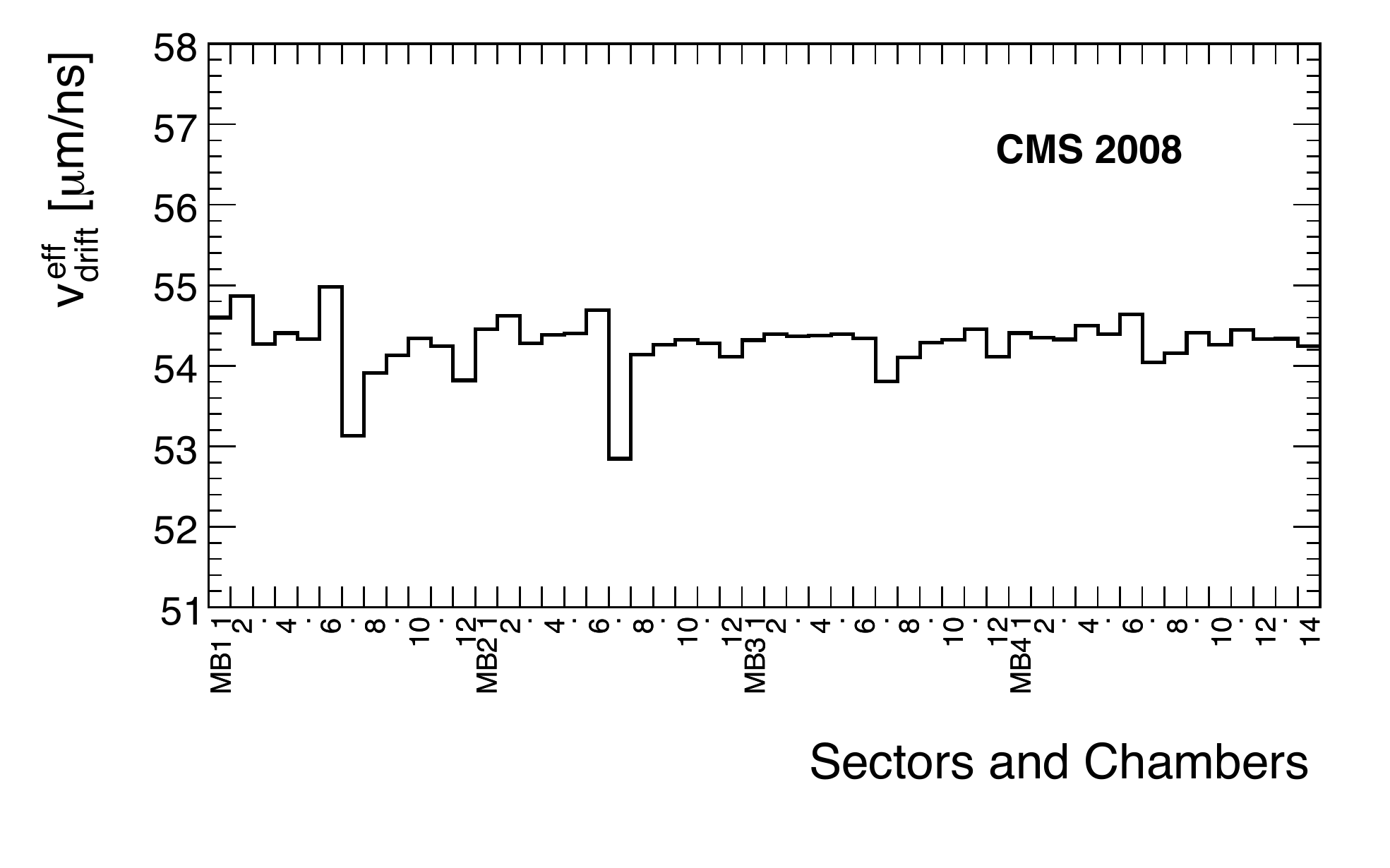}}
\end{tabular}
\caption{Drift velocities computed using the mean-time method for a
  run with $B=0~\rm T$ (top) and for a run with $B=3.8~\rm T$
  (bottom). Results are shown for the $r\phi$ super-layers of each
  chamber/sector of a representative wheel.  The other wheels show
  similar results.} \label{fig:vdrift}
 \end{center}
\end{figure}

Drift velocities measured for each chamber/sector and for a
representative wheel (other wheels give similar values) are shown in
Fig.~\ref{fig:vdrift} for two CRAFT runs, one without (top) and one
with (bottom) magnetic field.  The drift velocity has approximately a
constant value of $54.3~ \mu \rm m/ns$, although with some systematic
deviations, caused by limitations of the calibration procedure applied
to cosmic ray events.  As in the determination of the time pedestal,
these uncertainties originate mainly from the random arrival time of
cosmic muons relative to the clock cycle.

The $t_\textrm{trig}$ uncertainty of about $10~\rm ns$, seen in
Fig.~\ref{fig:Fig89a} (bottom), corresponds to a relative uncertainty
of about $2.5~\%$ on the drift velocity.  This is comparable to the
fluctuations observed in Fig.~\ref{fig:vdrift}, meaning that the
residuals calculated with the $v_\textrm{drift}^\textrm{eff}$ and
$t_\textrm{trig}$ constants do not represent a significant improvement
with respect to those shown in Fig.~\ref{fig:Fig1213}.

\begin{figure}[htbp]
  \begin{center}
   \resizebox{10cm}{!}{\includegraphics{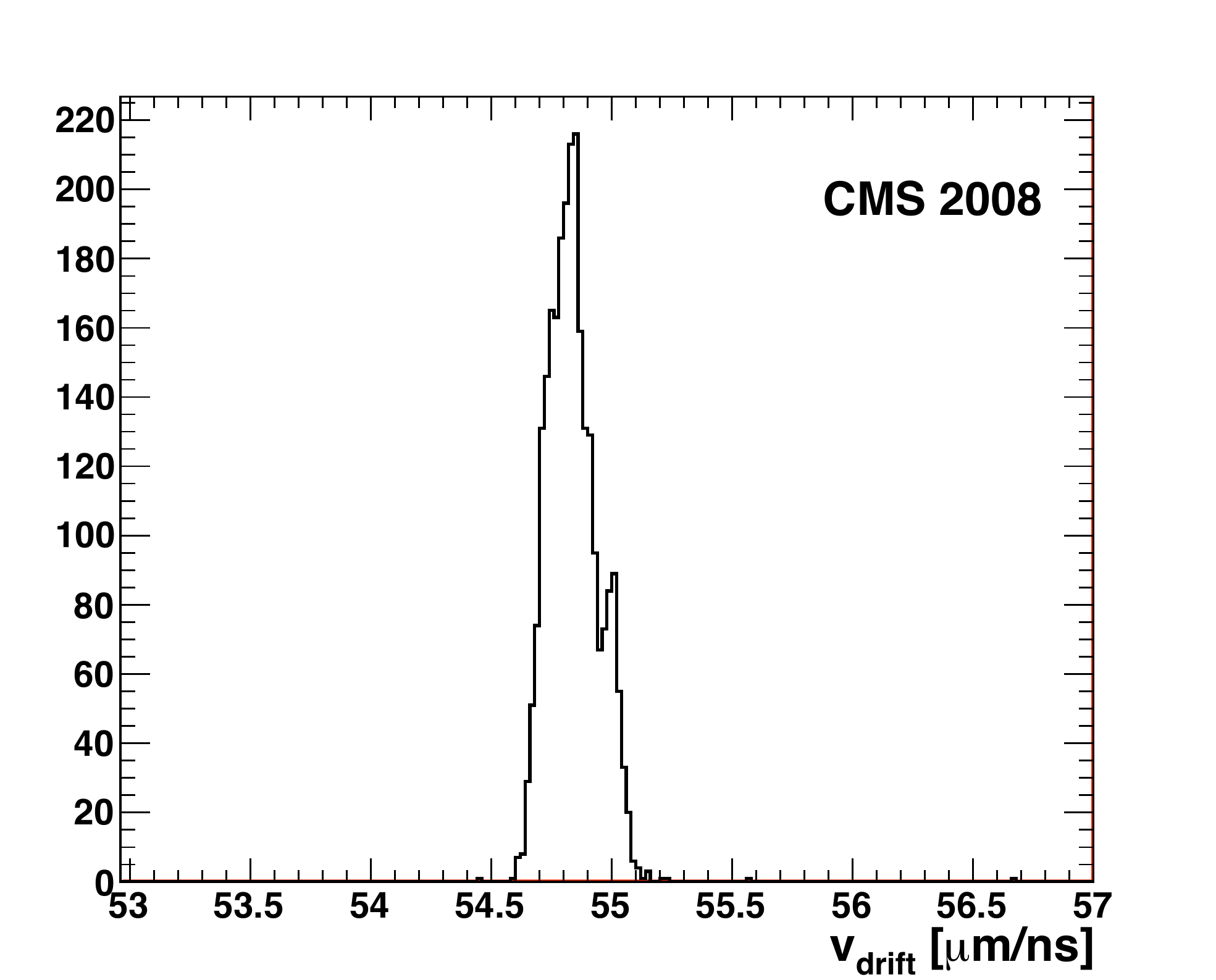}}
   \caption{Distribution of the drift velocity measured in one  Velocity Drift  
Chamber (VDC) during the entire data-taking period.}
   \label{fig:VDC}
  \end{center}
\end{figure}

The drift velocity distribution measured in one of the VDC chambers
(Section 3) 
during the full CRAFT period is shown in Fig.~\ref{fig:VDC}.
An average drift velocity value  of $54.8~ \mu \rm m/ns$ is observed, 
which  is slightly different ($0.5~\%$ higher) from the one obtained  
from the DT data, mainly because of the different
shape of the electric field in the DT drift cell.
The spread of the distribution is  better than $0.2~ \mu \rm m/ns$, and
shows that no major variations occurred in the gas mixture
or air contamination, during the entire data-taking period.

\section{The Calibration Workflow and the Monitoring of the Calibration Process}

A fast calibration of the DT system is vital in order to provide the
prompt data reconstruction with accurate calibration constants. The
number of calibration regions is a compromise between the need of
keeping things simple, not requiring too large event samples, and the need
of reducing systematic errors by separately calibrating regions where
parameters may have very different values. As mentioned in
previous sections, the super-layer granularity has been found to
be the most suitable calibration unit.

In order to reach the precision obtainable with the fast calibration,
about $10^4$ tracks crossing each super-layer are required.  During
LHC collision and cosmic ray data-taking periods, the calibration
parameters have to be produced, validated, and made available for use
in the reconstruction within one day of data-taking. 
However, after the start-up phase, it is
anticipated that at some point it will no longer be necessary to
update the DT calibrations on a daily basis; on the other hand, they
should be checked against a standard set in order to guarantee their
stability.

The workflow of the DT calibration has already been fully embedded into the 
central CMS calibration workflow at a very early stage. A more detailed 
description of the overall CMS calibration and alignment computing 
workflow used in the CRAFT exercise is given in   
Refs.~\cite{AlcaReco,AlcaRecoCRAFT}. 
Within the CMS calibration and alignment workflow, particular
selections of data, named \emph{AlCaReco}, were used. They 
contained a reduced number
of events and a reduced event content, providing the minimal
information to fulfill the requirements of the DT calibration
task. The sample is saved at the CERN Analysis Facility (CAF) and taken as input to the
calibration process.  The calibration algorithm runs at the CAF and
produces a set of constants, which undergoes a validation
procedure before being copied to the central CMS database,  where
they become available to the CMSSW offline software framework.

The DT calibration workflow has been used also during the
Computing, Software, and Analysis challenge (CSA08), described in
Ref.~\cite{AlcaReco}, which simulated with large event samples the conditions
expected at LHC startup. This exercise simulated the production rate
of the calibration conditions as it will happen during real collision
data-taking.  The long CRAFT data-taking period served as a thorough test of
this workflow with the real detector.

The quality and stability of the calibration constants is a crucial
part of the procedure and must be continuously monitored. 
Therefore, validation
procedures have been set up within the central CMS Data
Quality Monitoring (DQM) framework. A detailed description of the CMS
DQM structure is given in Ref.~\cite{DQM}.

Data quality assessment for the DT calibration constants consists
mainly in defining the acceptance criteria used to validate the
constants, in the monitoring of time stability, and in the checking
of continuous trends or sudden changes in operating conditions.  The
quality tests to assess the validation of the constants and to monitor
their time stability are applied to the residual distributions
calculated at the different steps of the calibration workflow.  The
comparison of the currently produced calibration constants with a
reference set gives an indication of the stability of each
particular calibration constant.

All the calibration constants described in the present paper have
their validation as well as monitoring process, and for each of them
detailed and summary DQM plots are provided.  The DT condition
constants have been monitored through the entire CRAFT data-taking
period and have shown generally a good stability in time.

\section{Drift Velocity Analysis}

The drift velocity obtained with the calibration procedure described
in Section 7 is derived from the measurements of the drift time and,
as already mentioned, is limited by the uncertainty on the arrival
time of the cosmic ray muons.

  \begin{figure}[htbp]
  \begin{center}
   \resizebox{10cm}{!}{\includegraphics{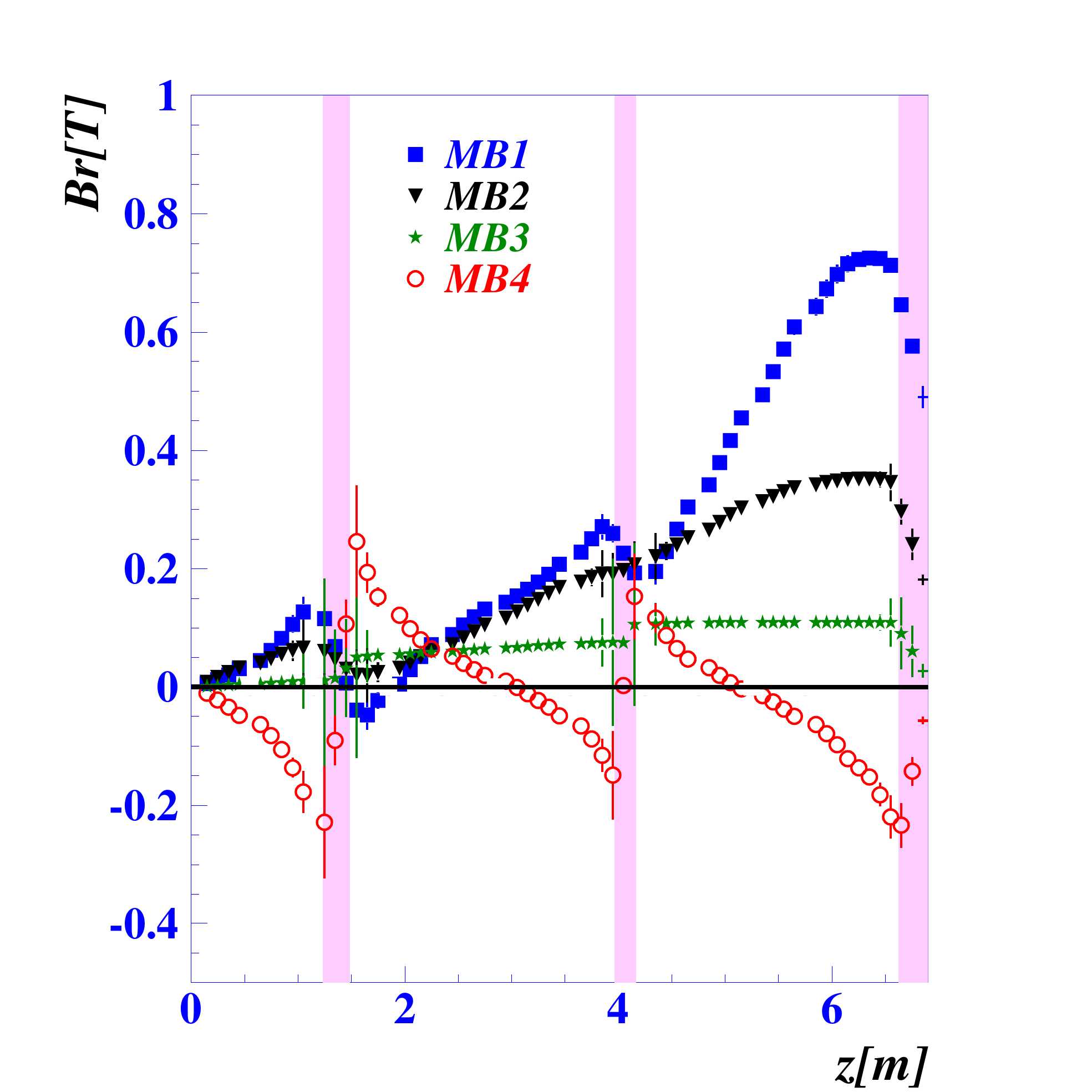}}
   \caption{Computed radial component of the
     magnetic field in the muon barrel chambers, for the different
     wheels, as a function of $z$.}
   \label{fig:P14}
  \end{center}
\end{figure}

A more detailed analysis of the drift velocity is presented in this
section, taking into account the precise 3D space-time relationship for
the hit reconstruction. In particular, it considers the influence of
the magnetic field as a function of the position along the wire.

The presence of a radial magnetic field distorts the drift lines of
the drifting electrons, because of the Lorentz force, resulting in a
variation of the effective drift velocity. Figure~\ref{fig:P14} shows
that the radial field component is not very high in the muon barrel
chambers, except in the MB1 chambers of the outer wheels, closest to
the endcaps.  In these regions, the radial component of the magnetic
field can be as high as $0.8~\rm T$, and changes significantly along
the $z$ axis, resulting in a variation of the effective drift velocity
along the wire of each single cell, for $r\phi$ super-layers. The
effect of the magnetic field has been studied in test beams with small
prototypes~\cite{vDriftComm1}, and more recently in the Magnet Test
and Cosmic Challenge (MTCC), using cosmic rays in the CMS surface
hall. These studies showed that the chambers maintain a good trigger
and event reconstruction functionality, even in the most critical
regions~\cite{t0seg}.

In the method presented in this section, a full reconstruction of the
trajectory within the muon system is performed to determine the drift
velocity.  In the first step, a pattern recognition algorithm is
applied to identify hits belonging to the same track.  Once the hits
have been identified, the track is reconstructed under the assumption
of a $54.3~\mu \rm m/ns$ nominal drift velocity.  In the second step
the track is refit treating as free parameters the drift velocities at
each hit and the
time of passage of the muon through the chamber.  The method is
applied to the $r\phi$ view of the track segment in one chamber, where
there are eight measured points in most cases. The $z$ super-layers,
where only four points are available, at most, are less significant
for this analysis.  The drift velocity is taken to be the mean
value of the track-by-track drift-velocity distribution.

  \begin{figure}[htbp]
  \begin{center}
   \resizebox{10cm}{!}{\includegraphics{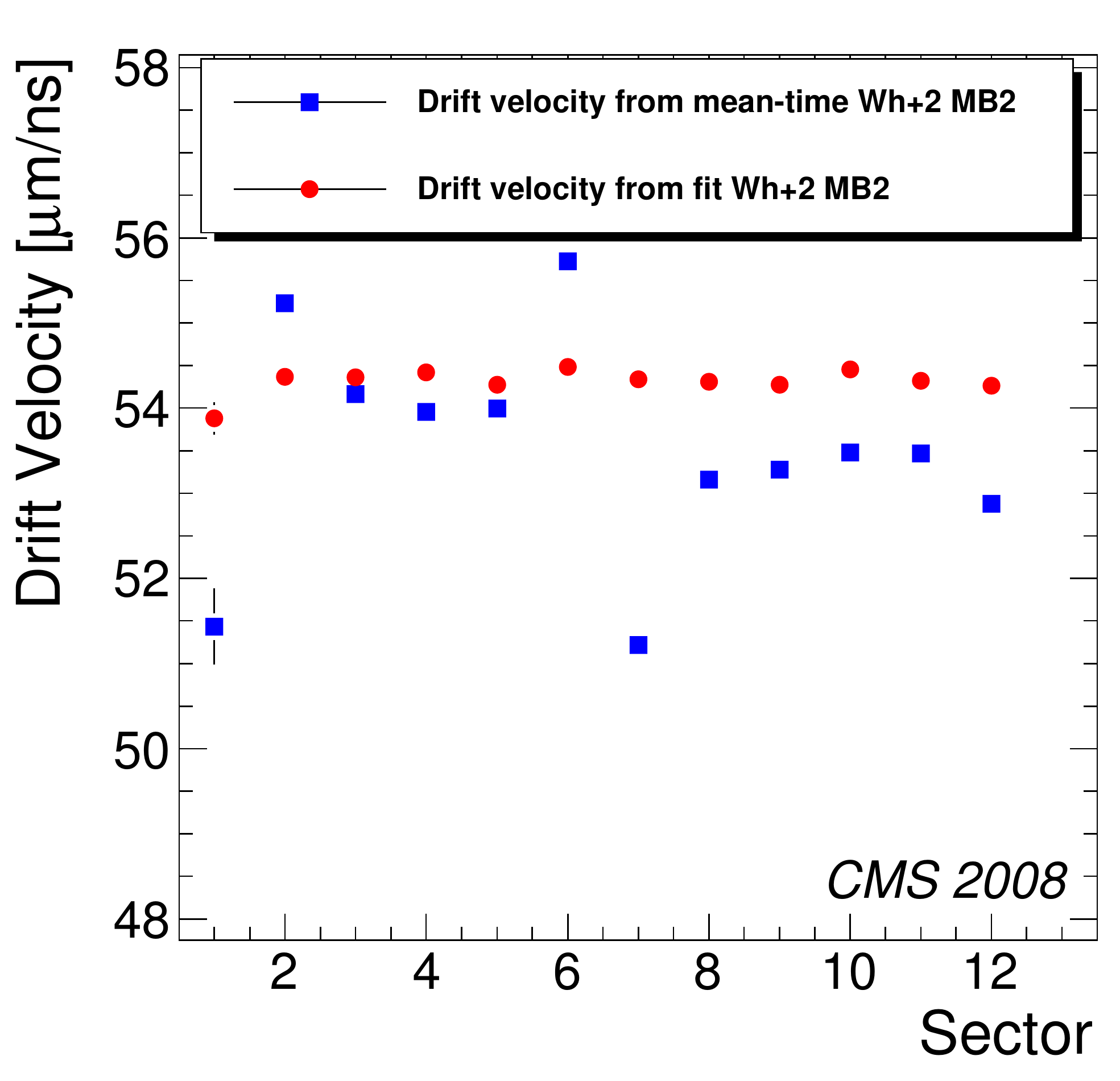}}
   \caption{Mean values of the drift velocity for the MB2 chambers of
     Wheel +2, using the mean-time (squares) and fit (circles) methods.
     The differences between sectors when using the mean-time method
     are due to $t_\textrm{trig}$ uncertainties that are not present in the
     fit method.}
   \label{fig:Fig14}
  \end{center}
\end{figure}

Figure~\ref{fig:Fig14} shows the mean values of the drift velocity for
the MB2 chambers of Wheel +2, using the mean-time method described in
Section 7 and the fit method described here.  When using the mean-time
method, the drift velocities have large systematic fluctuations from
one sector to the other. This is related to the errors on the
$t_\textrm{trig}$ determination described in Section 6, which 
cancel when the fit method is used.

\begin{figure}[htbp]
  \centering
  \begin{minipage}[c]{.45\textwidth}
     \includegraphics[width=1.\textwidth]{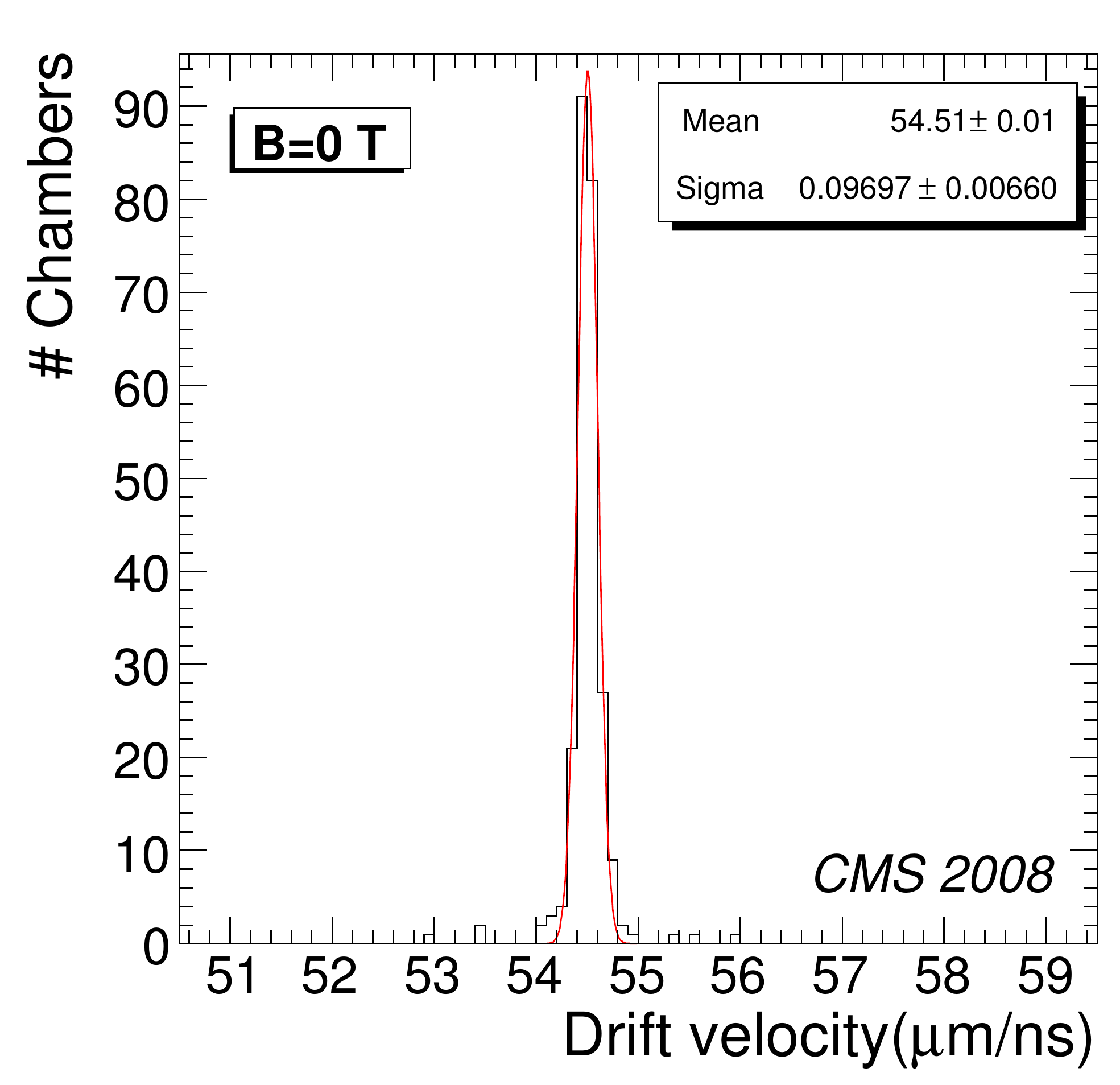} 
  \end{minipage}
  \hspace{10mm}
  \begin{minipage}[c]{.45\textwidth}
    \includegraphics[width=1.\textwidth]{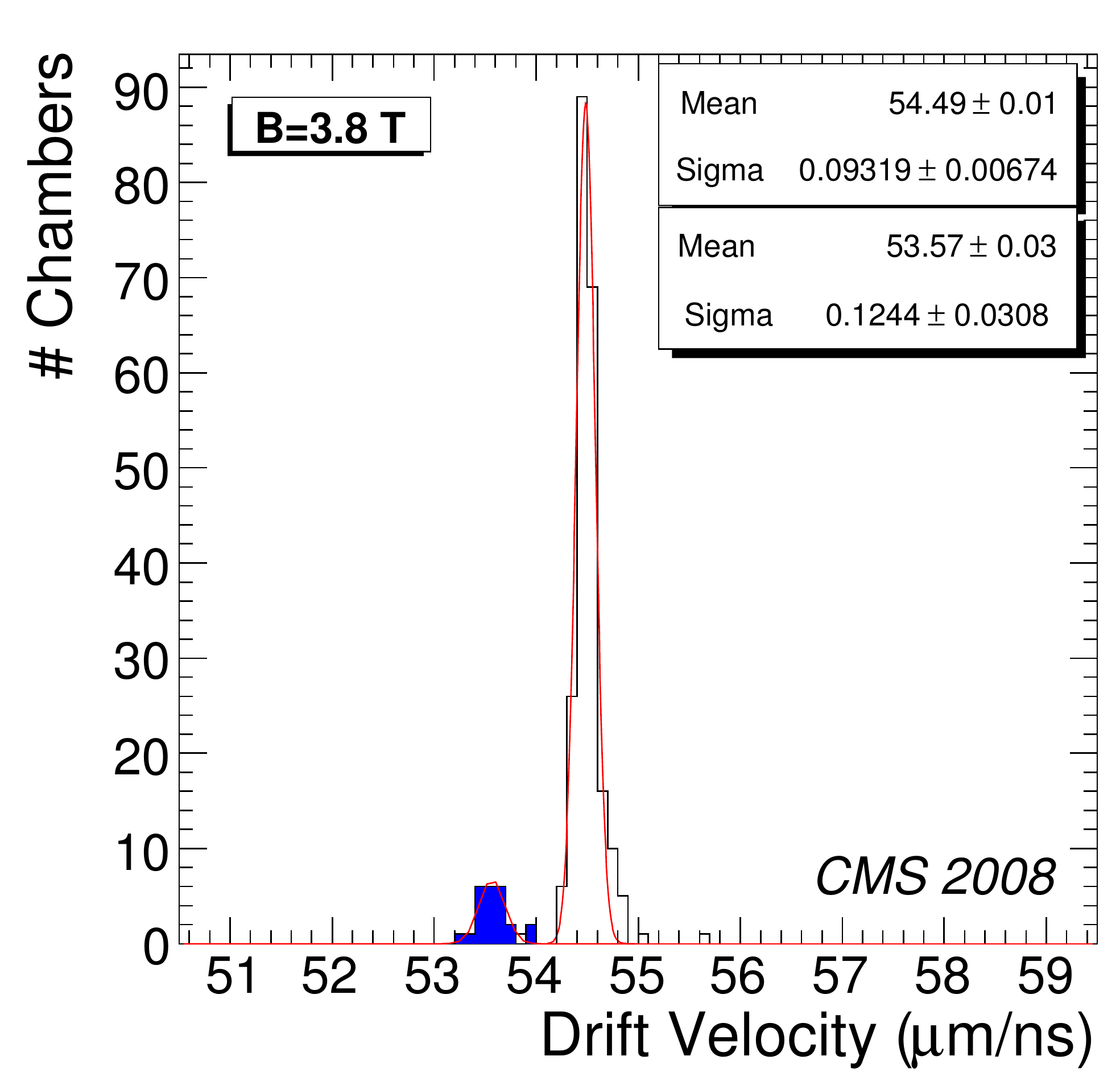}
  \end{minipage}
  \caption{Drift velocities for $B=0~\rm T$ (left) and $B=3.8~\rm T$
    (right).  The small peak on the right panel corresponds to the MB1
    chambers of Wheel +2 and Wheel $-$2, and shows the influence of a
    higher magnetic field in these regions.\label{fig:VDAll}}
\end{figure}

The average drift-velocity values from the fit method, for all the
chambers, are shown in Fig.~\ref{fig:VDAll}, for runs without and with
magnetic field (of $3.8~\rm T$).

The data at $B=0~\rm T$ show an average value of $54.5~ \mu \rm m/ns$
for the drift velocity and a standard deviation indicating that
differences between chambers are in the order of $0.2~\%$. For
$B=3.8~\rm T$, a second peak is observed at $53.6~\mu \rm m/ns$. This peak
corresponds to the MB1 chambers of the external wheels (Wheel +2 and
Wheel $-$2) and is due to the presence of a higher radial magnetic
field.

Similar values of the drift velocity have been obtained using the same
calibration procedure applied to the simulated pp collision data.
These results, presented in Ref.~\cite{CSA08}, indicate that the
calibration algorithm delivers a more uniform response in the case of
collision data and that a large fraction of the fluctuations observed
in the drift velocity calibration from CRAFT data may be attributed to
the topology and timing of the cosmic ray events.

   \begin{figure}[htbp]
  \begin{center} 
   \resizebox{17cm}{!}{\includegraphics{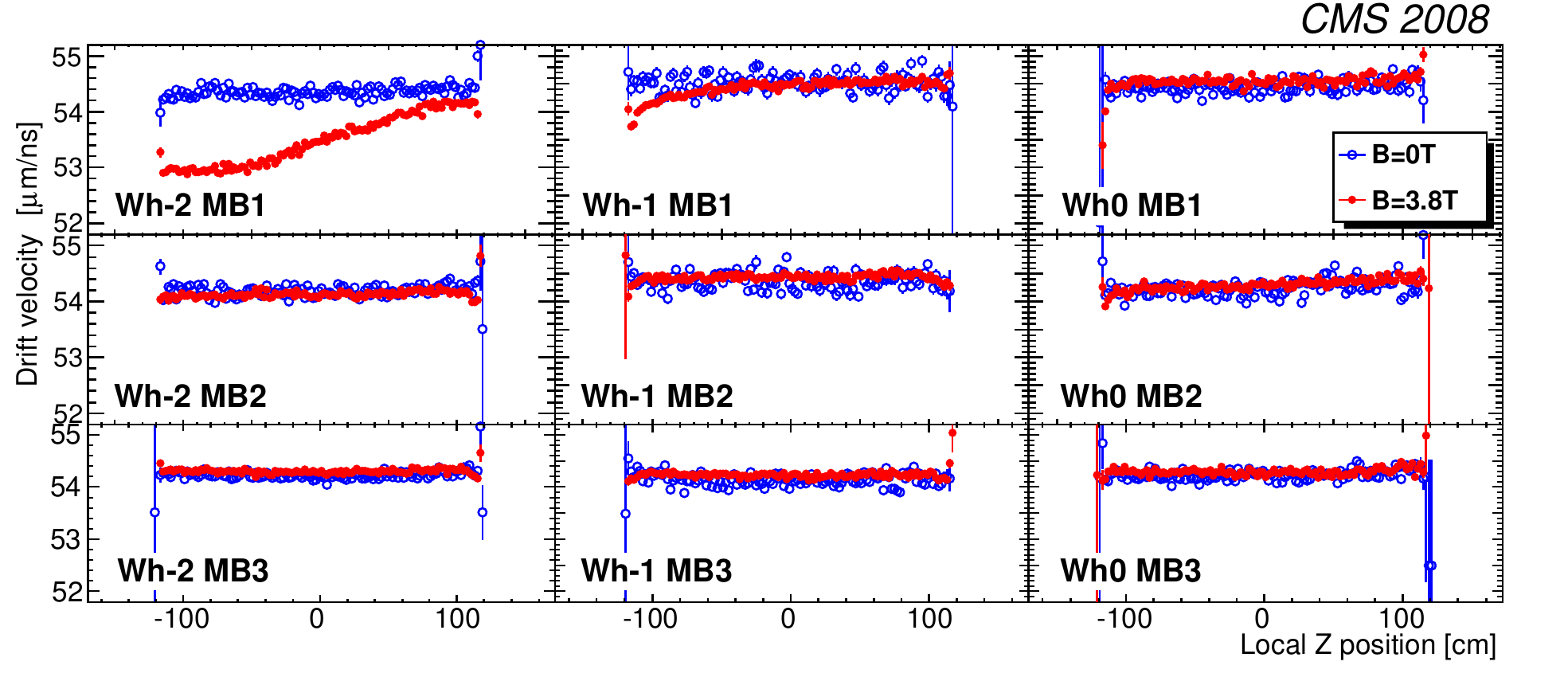}}
   \caption{Drift velocities calculated using the muon track fit
     method described in this section. The values are shown as a
     function of the $z$ position (measured by the $z$ super-layers),
     and for $B=0~\rm T$ and $B=3.8~\rm T$.}
   \label{fig:Fig16}
  \end{center}
\end{figure}

The effect on the drift velocity of the variation of the radial
magnetic field along the $z$ coordinate is shown in
Fig.~\ref{fig:Fig16}, as calculated with the fit method. Positive
wheels (+1 and +2) are not in the figure but show the same behavior as
their symmetric wheels ($-$1 and $-$2, respectively).

The presence of the radial component of the magnetic field affects, as
expected, only the MB1 chambers, primarily in the external wheels but
some effects are also observed in Wheels +1 and $-$1. The variation
along $z$ for the MB1 chambers of Wheels +2 and $-$2 is below $3~\%$,
less than expected from the MTCC results~\cite{t0seg}, after taking
into account the differences of the magnetic field conditions between
both periods ($B=4~\rm T$ during the MTCC in the surface hall,
$B=3.8~\rm T$ in the underground experimental hall).

This analysis of the drift velocity is very sensitive to the field
strength and, in fact, provided the first evidence of a systematic
deviation from the true field strength in the field map. 
 A new detailed
calculation of the magnetic field has been performed~\cite{NotaNic}, and the new
analyses, currently in progress, show a $z$ dependence in satisfactory
agreement with the expectations.

\begin{figure}[htbp]
\begin{center}
\begin{tabular}{c}
\resizebox{10cm}{!}{\includegraphics{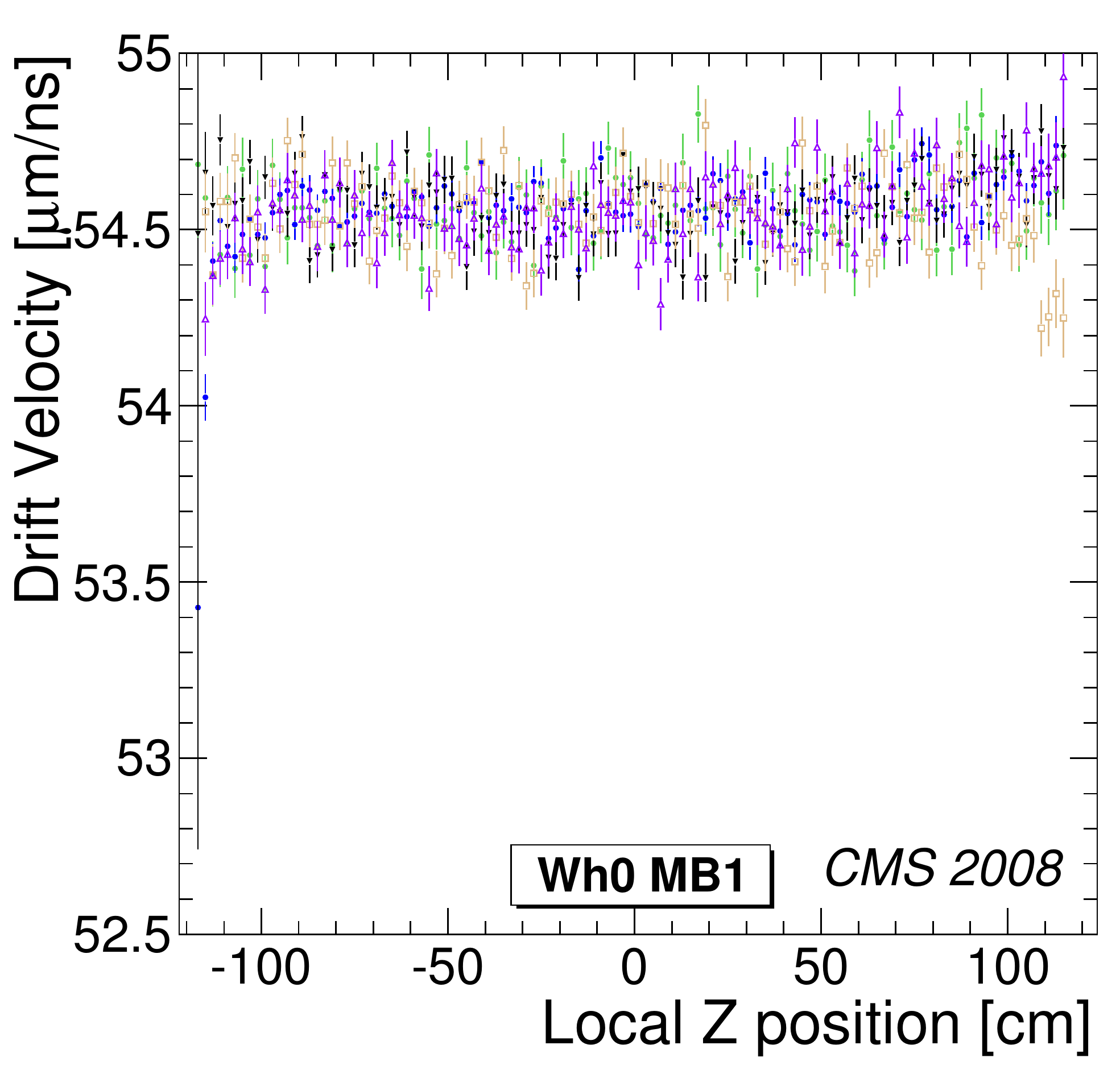}} \\
\resizebox{10cm}{!}{\includegraphics{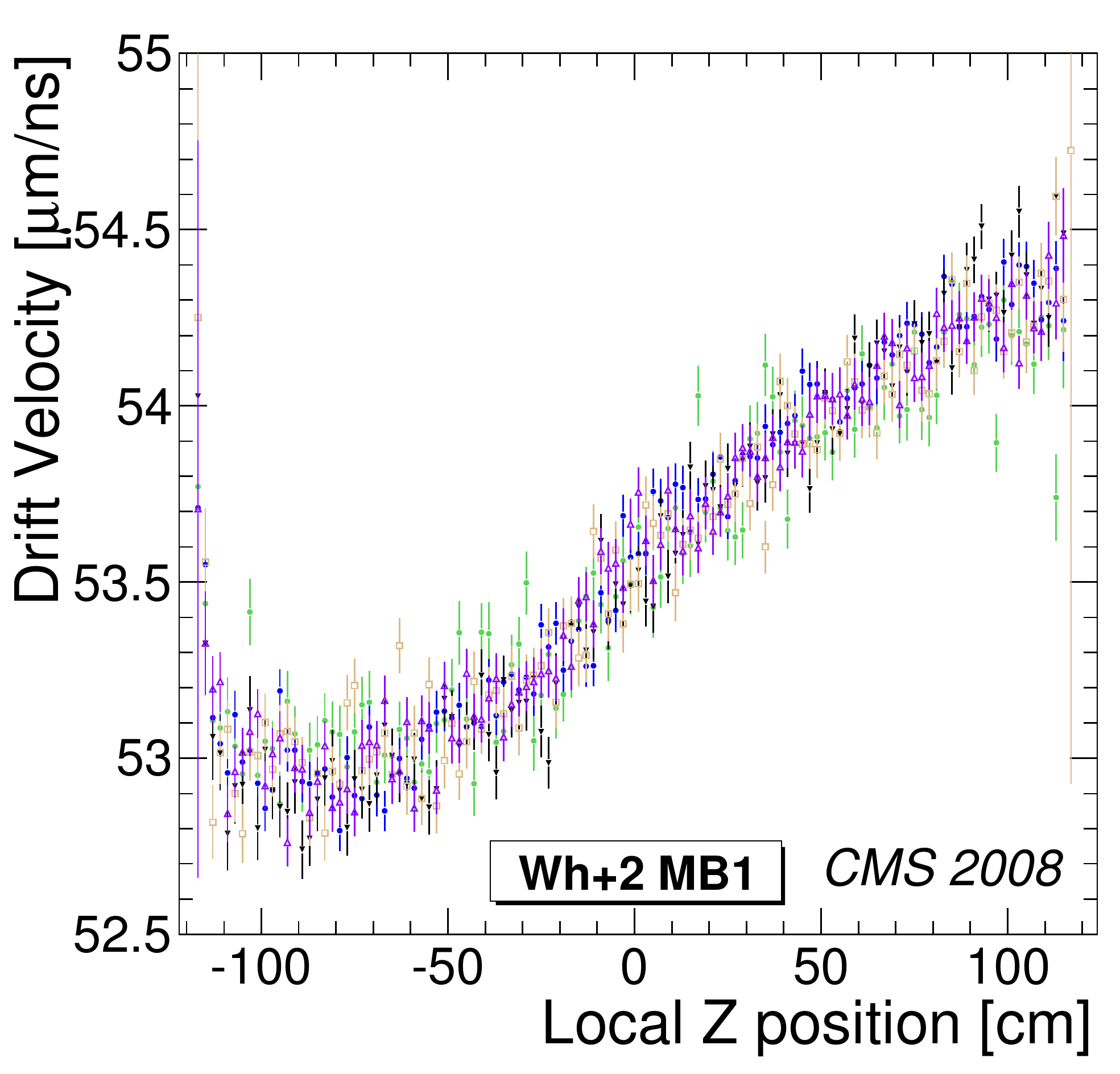}} \\
\end{tabular}
\caption{Drift velocities as a function of the local $z$ position for
  MB1 chambers of some representative sectors of Wheel 0 (top) and
  Wheel +2 (bottom).  Different sectors are  indicated by different
  grey tones.}
\label{fig:VDvsZ_MB1}
\end{center}
\end{figure}

All sectors in the same wheel show the same behavior, 
as illustrated in Fig.~\ref{fig:VDvsZ_MB1}, 
where the values of the drift velocity along the z axis are shown for
the MB1 chambers of some representative sectors of Wheels +2 and 0. 

The drift velocity calculation, performed in this section, 
provides a better spatial resolution of the chambers with respect to 
the one obtained in Section 7. This improvement is obtained with an 
extended track fit method which determines the drift velocity 
and the time of passage of the muon simultaneously with 
the regular track parameters. The detailed analysis of the spatial 
resolution for the cosmic ray data taking in 2008 is given in
Ref.~\cite{NotaUgo}. The value obtained is about  $250~\mu \rm m$, 
in fair agreement with the requirements for collision dataa~\cite{MUON}. 
%A consequence of the more precise drift velocity calculation, obtained
%with the track fit method, is of a better spatial
%resolution of the chambers~\cite{NotaUgo}.  The value obtained is about $250~\mu \rm
%m$, in fair agreement with the requirements for collision
%data~\cite{MUON}.

\section{Summary}

This paper describes the calibration of the CMS Drift-Tubes system and
presents results from the cosmic ray data-taking period which took
place in 2008.

The complete calibration workflow has been applied to the data.  It
performed efficiently, monitoring the stability of the produced
constants, and delivering with very low latency the calibration
constants to the conditions database used by the offline
reconstruction.

The first calibration step is the identification and masking of noisy
channels to have a clean structure of the drift time distribution. The
fraction of noisy cells was stable and about $0.01~\%$. The average
noise rate was $\sim 4~\rm Hz$.

The time pedestals, after having been corrected for the inter-channel
synchronization, noisy channels, and the time of flight between upper
and lower sectors, show a constant behavior in the entire DT system.
Due to the particular topology of the cosmic ray events, the time
pedestals are poorly defined for the sectors with chambers in the
vertical plane, where cosmic ray tracks with large impact angles are
measured. For all the other sectors, an uncertainty of the order of
$10~\rm ns$ is observed. This value agrees with the uncertainty of the
arrival time of cosmic ray muons within the clock cycle.

The drift velocity calibration results show an approximately constant
value of $54.3~\mu \rm m/ns$ for all the chambers of the DT system,
with a relative systematic uncertainty of $2.5~\%$.  This uncertainty
originates from the measured drift time, used in the mean-time method,
which is limited by the uncertainty of the arrival time of cosmic ray
muons.  
This explains why the obtained spatial resolution is worse than 
would be expected with collision data. 

A more refined analysis of the drift velocity has been performed,
exploiting the full potential of the CMS offline software for data
reconstruction.  It uses a track fitting procedure which leaves as
free parameters the drift velocity and the time of passage of the
muons through the chambers.  Cosmic ray data with and without magnetic
field have been studied.  Without magnetic field, a constant average
value of $54.5 ~\mu \rm m/ns$ has been observed, with an error of
$0.2~\%$; when the field strength is $3.8~\rm T$, the innermost
chambers of the external barrel wheels measure a lower value, as
expected, of about $53.6~\mu \rm m/ns$.  These results confirm what
was observed in an analysis performed on simulated collision data and
provide a spatial resolution that is close to the design performance.

\section*{Acknowledgements}

We thank the technical and administrative staff at CERN and other CMS
Institutes, and acknowledge support from: FMSR (Austria); FNRS and FWO
(Belgium); CNPq, CAPES, FAPERJ, and FAPESP (Brazil); MES (Bulgaria);
CERN; CAS, MoST, and NSFC (China); COLCIENCIAS (Colombia); MSES
(Croatia); RPF (Cyprus); Academy of Sciences and NICPB (Estonia);
Academy of Finland, ME, and HIP (Finland); CEA and CNRS/IN2P3
(France); BMBF, DFG, and HGF (Germany); GSRT (Greece); OTKA and NKTH
(Hungary); DAE and DST (India); IPM (Iran); SFI (Ireland); INFN
(Italy); NRF (Korea); LAS (Lithuania); CINVESTAV, CONACYT, SEP, and
UASLP-FAI (Mexico); PAEC (Pakistan); SCSR (Poland); FCT (Portugal);
JINR (Armenia, Belarus, Georgia, Ukraine, Uzbekistan); MST and MAE
(Russia); MSTDS (Serbia); MICINN and CPAN (Spain); Swiss Funding
Agencies (Switzerland); NSC (Taipei); TUBITAK and TAEK (Turkey); STFC
(United Kingdom); DOE and NSF (USA). Individuals have received support
from the Marie-Curie IEF program (European Union); the Leventis
Foundation; the A. P. Sloan Foundation; and the Alexander von Humboldt
Foundation.

\bibliography{auto_generated}   % will be created by the tdr script.
\cleardoublepage\appendix\section{The CMS Collaboration \label{app:collab}}\begin{sloppypar}\hyphenpenalty=500\textbf{Yerevan Physics Institute,  Yerevan,  Armenia}\\*[0pt]
S.~Chatrchyan, V.~Khachatryan, A.M.~Sirunyan
\vskip\cmsinstskip
\textbf{Institut f\"{u}r Hochenergiephysik der OeAW,  Wien,  Austria}\\*[0pt]
W.~Adam, B.~Arnold, H.~Bergauer, T.~Bergauer, M.~Dragicevic, M.~Eichberger, J.~Er\"{o}
% ...
\vskip\cmsinstskip
% ...

\end{sloppypar}
\end{document}